\documentclass[%
reprint,
superscriptaddress,
showpacs,
 amsmath,amssymb,
 aps,
prb,
twocolumn,
oneside,
floatfix,
]{revtex4-1}

\usepackage[lined,boxed,commentsnumbered]{algorithm2e}
\usepackage{graphicx}
\usepackage{dcolumn}
\usepackage{bm}
\usepackage[utf8]{inputenc}
\usepackage{amssymb,bbm}
   
\usepackage{braket}
\usepackage{xcolor}
\usepackage{color}
\usepackage{soul}

\newcommand{\si}{\sigma}
\newcommand{\eps}{\epsilon}
\newcommand{\dt}{\tilde{d}}
\newcommand{\pt}{\tilde{p}}
\newcommand{\rot}{e^{i\theta}}
\newcommand{\mrot}{e^{-i\theta}}
\newcommand{\+}{^{\dagger}}

\newcommand{\Tr}{\mathrm{Tr}}
\newcommand{\intd}{\mathrm{d}}

\newcommand{\iintbeta}{\iint_{0}^{\beta}\mathrm{d}\tau\mathrm{d}\tau'}

\newcommand{\upa}{\uparrow}
\newcommand{\doa}{\downarrow}

\begin{document}

\preprint{APS/123-QED}

\title{Intershell Interactions in Correlated Materials: A Slave Rotor Approach}

\author{Jakob Steinbauer}
\email{jakob.steinbauer@polytechnique.edu}
\affiliation{CPHT, CNRS, Ecole Polytechnique, Institut Polytechnique de Paris, F-91128 Palaiseau, France}
\author{Silke Biermann}%
\affiliation{CPHT, CNRS, Ecole Polytechnique, Institut Polytechnique de Paris, F-91128 Palaiseau, France}
\affiliation{Coll\`{e}ge de France, 11 place Marcelin Berthelot, 75005 Paris, France}
\affiliation{Department of Physics, Division of Mathematical Physics, Lund University, Professorsgatan 1, 22363 Lund, Sweden}
\affiliation{European Theoretical Spectroscopy Facility, Europe}
\date{\today}

\begin{abstract}
The treatment of intershell interactions remains a major challenge in the theoretical description of strongly correlated materials. 
Most previous approaches considered the influence of intershell interactions at best in a static fashion, neglecting dynamic effects. In this work, we propose a slave-rotor method that goes beyond this approximation by incorporating the effect of intershell interactions in a dynamic manner. Our method is derived and implemented as a quantum impurity solver in the context of \emph{dynamical mean field theory} and benchmarked on a two-orbital model system. The results from our slave-rotor technique are found to be in good agreement with our reference calculations that include intershell interactions explicitly. We identify and analyze qualitative features emerging from the dynamic treatment. Our results thus provide qualitatively new insights, revealing the ambivalent effect of intershell interactions in strongly correlated materials.
  
\end{abstract}

\pacs{Valid PACS appear here}
\maketitle


\section{\label{sec:Introduction} Introduction}



Transition metal oxides (TMOs) constitute one of the most fascinating classes of solids. Historically, they were the first systems for which a breakdown of band theory was observed\cite{Boer_1937,Mott_1937} and since then, the interest in these compounds has hardly declined.
The interaction-driven localization of electrons\cite{BrinkmanRice,MIT_DMFT_RozenbergKotliar} -- the Mott metal-insulator transition and
various ordering phenomena involving spin-, charge-, or orbital degrees
of freedom
or superconductivity
are only examples of the rich physics displayed
in these compounds \cite{ImadaReview}.\par

From a microscopic perspective, the properties of TMOs are largely
dictated by the Coulomb interactions felt by the electrons in the
$d$-orbitals on the transition metal ions. Even though these interactions
are screened by intra-atomic, anionic and inter-site processes, typical
values of the effective interactions still exceed several eV, and are
thus comparable to or even larger than the relevant kinetic energies
\cite{Hubbard_model,Gunnarsson1,Gunnarsson_Anisimov,Anisimov_cLDA, cRPA_Aryasetiawan1,cRPA_Aryasetiawan,
  Vaugier_PRB2012,Panda_PRB,VanRoekeghem_PRB2016,Hansmann_PRL2013}. 

\par
Early theories, developed since the 1960s, have set up
effective lattice models for describing the physics emerging from the
interplay of Coulomb interactions and kinetic energy in TMOs.
The most famous of these is probably the Hubbard model\cite{Hubbard_model}
which offers a minimal description of correlated systems, considering
only a single orbital. Despite its simplicity, it incorporates many of
the effects found in strongly correlated materials, such as the
the Mott metal-insulator transition\cite{BrinkmanRice,
  MIT_DMFT_RozenbergKotliar},
antiferromagnetism\cite{hubbard_antiferromagnet_hirsch, hubbard_antiferromagnet_vitaly} or -- depending on the lattice under consideration --
superconductivity\cite{scalapino_hubbard,tJ_rice_troyer,hubbard_absence_of_SC,spin_freezing_cuprates}.\par

Based on these early successes, more recently, intense research lines
focusing on a quantitative theoretical description of correlated TMOs
have emerged. 
The single orbital Hubbard model is indeed often a crude oversimplification.
In most materials, more than one electronic band crosses the Fermi level
and a faithful description of the physics thus requires low-energy models
comprising multiple orbitals. In the case of TMOs, many of their interesting
electronic response properties can be attributed to the strongly
correlated nature of the partially filled transition metal d-shells.
Orbital degeneracies (or lifting thereof by crystal field splittings)
\cite{Martins1,Martins_2017,Poteryaev},
electronic filling and hybridizations provide additional degrees of freedom
contributing to the rich landscape of phenomena observed in TMOs\cite{ImadaReview}.

The challenge to theory consists in the derivation of effective
low-energy Hamiltonians taking these microscopic details into account.
Ideally, this should be done by means of appropriate downfolding techniques \cite{Hirayama} applied directly to the continuum Coulomb Hamiltonian.
However, constructing the interacting part of a low-energy Hamiltonian
is a delicate issue. Here, immense progress has been made with the development of the \emph{constrained random phase approximation}\cite{cRPA_Aryasetiawan1,cRPA_Aryasetiawan} (cRPA) method. Within the cRPA, the partially screened effective Coulomb interactions are calculated by considering screening from degrees of freedom that lie outside of the low-energy window of interest. \par

Such (semi-)quantitative approaches nowadays provide access to explicit
estimates of the effective Coulomb interactions 
and have revealed that not only the celebrated local ``Hubbard U''
on the transition metal $d$-shell is non-negligible: Depending on the
compound, non-local interactions \cite{} or intershell interactions
between the $d$-shell and ligand $p$-orbitals can become important as well
\cite{}. This is consistent with early discussions on screening
phenomena in solids \cite{brandow_dp,perfect_screening_herring,FujimoriMinami}
and has also been the
starting point to recent investigations studying the renormalization
of local Coulomb interactions by intershell interactions \cite{shell_folding}.

Including the ligand p-states can have important consequences on the physics\cite{dp_millis,d_dp_parragh}, especially for compounds assigned to the family of the late TMOs, where charge transfer is found to play an important role\cite{ZaanenSawatzkyAllenPRL,Sawatzky_NIO_cluster}. Zaanen, Sawatzky and Allen early on argued that besides the effective d orbital Coulomb interaction, the charge-transfer energy constitutes a central quantity in the characterization of transition metal oxides.\cite{ZaanenSawatzkyAllenPRL} \par 
To describe the physics of the late TMOs, the state-of-the-art thus consists in considering so-called d-p models that include both transition metal d orbitals, as well as ligand p orbitals.
To get on top of the increasing complexity of performing numerical calculations on such models, the p-states are commonly treated as uncorrelated and intershell interactions are either ignored or treated within static mean field schemes. 

Recently, an approach to include the effect of intershell interactions was suggested with the so-called \emph{shell-folding}\cite{shell_folding} scheme. Within \emph{shell-folding}, interactions of the correlated orbitals are treated on the same footing as intershell interactions with the ligands. It's basic assumption is ``perfect screening'' of the correlated orbitals by the neighboring ligands, i.e. that the total charge within each Wigner-Seitz cell is a conserved quantity\cite{perfect_screening_herring}. This permits to derive an effective low-energy model without intershell interactions and with statically renormalized (reduced) effective Coulomb interactions.\par
However elegant the shell-folding approximation is, its dependence on the ``perfect screening'' assumption means that it will break down whenever charge fluctuations become too strong. Instead of pursuing such a procedure, one might consider skipping the final down-folding step and directly attempt to solve the low-energy model including intershell correlations and correlated ligands. This might work for the most simple model Hamiltonians; in the case of more realistic multi-orbital systems, however, the problem quickly becomes intractable.\par
In this paper, we propose a novel method that goes beyond the shell-folding approximation by explicitly incorporating charge fluctuations at moderate computational cost. Our method is to be understood as a solver for impurity models with intershell interactions. A slave rotor\cite{SR_impurity,SR_meanfield} approach allows deriving a shell-folding-like Hamiltonian that can be treated by means of standard quantum Monte Carlo methods, while the  intershell contribution is treated analytically. A similar hybrid approach has already successfully been applied to models with dynamic Hubbard interaction\cite{SR_Krivenko}. \par
The paper is outlined as follows. 
In Section \ref{sec:model_intro} we describe the model systems under consideration and give a short review of the shell folding approximation upon which our technique is built. In Section \ref{sec:slave_rotor_derivation} we then derive our method to efficiently include dynamic effects from ligand interactions. We discuss approximations to be applied in practice, their physical meaning and possibilities for improvement.
In Section \ref{sec:Testing}, we test our slave rotor technique on a two-orbital model that incorporates intershell interactions in a minimal fashion. Our method is benchmarked in various parameter regimes and compared to the shell-folding technique, as well as calculations using a Hartree-Fock mean-field approximation.
In Section \ref{sec:discussion}, we summarize our findings and give an outlook of possible applications. 

\section{The Model: From the lattice to the impurity}\label{sec:model_intro}
As a starting point for our derivation, we consider a low energy model, comprised of orbitals from two different shells. Throughout the paper, we will refer to them as ``d'' and ``p'', inspired by the transition metal d orbitals and ligands p orbitals. We emphasize, however, that our scheme can be applied to a larger class of compounds, such as e.g. f-electron systems. The problem shall then be solved within \emph{dynamical mean field theory}\cite{DMFTkotliar_georges}, which maps the original lattice model onto an interacting impurity connected to an uncorrelated bath, which is to be determined self-consistently. \par
In the following sections, we will elaborate a method to effectively describe the d-p interactions of such models at the cost of solving a model with interacting d orbitals only. In subsection \ref{sec:SR_plus_DMFT}, we then discuss how this scheme can be used in the context of DMFT.

\subsection{The d-p model}
Within DMFT a multi-orbital lattice model is mapped onto a quantum impurity model, which is described by the following Hamiltonian
\begin{align}\label{eq:the_action}
H = H_{0} + H_{int} + H_{hyb} \text{ .} 
\end{align}
Here, we consider the case of a d-p model with $N^{d}+N^{p}$ 
orbitals, where $N^{d}$ and $N^{p}$ are the numbers of the correlated d orbitals and the ligand p orbitals, respectively.
The non-interacting part of the Hamiltonian then reads
\begin{align}\label{eq:action_0}
\begin{split}
H_{0}&=\sum_{m\si} d\+_{m\si}(-\mu+\eps_{m}^{d})d_{m\si}  \\
 &+ \sum_{n\si} p\+_{n\si}(-\mu+\eps_{n}^{p})p_{n\si}  \\
 & + \sum_{m n\si} \left( V_{mn}d\+_{m\si}p_{n\si} + h.c.\right) \text{ ,}
 \end{split}
 \end{align}
 and explicitly considers on-site hybridization between the orbitals of the two shells. The second term in \eqref{eq:the_action} describes the
Coulomb interactions  

 \begin{align}\label{eq:action_int}
 \begin{split}
H_{int} &= \frac{1}{2} \sum_{(m\si)\neq(m'\si')}U^{dd}_{m\si m'\si'}\hat{n}^{d}_{m\si}\hat{n}^{d}_{m'\si'}\\
&+ \frac{1}{2}\sum_{(n\si)\neq(n'\si')} U^{pp}_{n\si n'\si'}\hat{n}^{p}_{n\si}\hat{n}^{p}_{n'\si'} +  U^{dp} \sum_{\si \si'}\hat{N}^{d}_{\si}\hat{N}^{p}_{\si'} \text{ .} \end{split} 
\end{align}
It includes intra- and intershell interactions, with $\hat{N}^{(d/p)}_{\si} = \sum_{m} \hat{n}^{(d/p)}_{m\si}$.\\
Finally, the last term in \eqref{eq:the_action} describes the hybridization between the correlated quantum impurity and the surrounding bath sites. It takes the form {\color{black}
\begin{align}\label{eq:Ham_hyb}
\begin{split}
H_{hyb} &=   \sum_{mm'k\si} d\+_{m\si} V^{dd}_{mm'k} b^{d}_{m'k\si} + h.c\\
&+ \sum_{m n k\si} \left\{d\+_{m\si}V^{dp}_{mn k} b^{p}_{nk\si} + p\+_{n\si}V^{pd}_{nm k} b^{d}_{mk\si}\right\}  +h.c.\\
&+ \sum_{nn' k\si} p\+_{n\si}V^{pp}_{nn'k}b^{p}_{n'k\si} + h.c.\\
&+ \sum_{mk\si} E^{d}_{mk}b^{d\dagger}_{mk\si}b^{d}_{mk\si} + \sum_{nk\si}E^{p}_{nk}b^{p\dagger}_{nk\si} b^{p}_{nk\si}\text{ ,}
\end{split}
\end{align}}
where the $V$'s describe the hybridization strengths, while the $E$'s specify the bath-site energies.\par
Switching to a functional integral formalism, this last term can be brought to a more elegant form by integrating out the bath sites
 \begin{align}\label{eq:action_hyb}
 \begin{split}
 S_{hyb} = \sum_{i\omega} &\left\{ \sum_{mm'\si} d_{m\si}^{\dagger}(i\omega) \Delta^{dd}_{mm'}(i\omega) d_{m'\si}(i\omega) \right.\\
+& \left.\sum_{nn'\si} p_{n\si}^{\dagger}(i\omega) \Delta^{pp}_{nn'}(i\omega) p_{n'\si}(i\omega) \right.\\
+&\left. \sum_{mn\si} \left( d\+_{m\si}(i\omega) \Delta^{dp}_{mn}(i\omega) p_{n\si}(i\omega) + h.c. \right) \right\} \\
=\sum_{i\omega\sigma} &\begin{pmatrix}
\vec{d}\+_{\si}(i\omega) &  \vec{p}\+_{\si}(i\omega)
\end{pmatrix} \mathbf{\Delta}^{dp}(i\omega) \begin{pmatrix}
\vec{d}_{\si}(i\omega)\\
\vec{p}_{\si}(i\omega)
\end{pmatrix}\text{ ,}
\end{split}
\end{align}
where in the last line, we introduced a vector/matrix notation with
\begin{align}\label{eq:dp_compact}
\begin{pmatrix}
\vec{d}\+_{\si} &  \vec{p}\+_{\si}
\end{pmatrix} = \begin{pmatrix}
d\+_{1\si} & d\+_{2\si} &\dots & d\+_{N^{d}\si} & \vec{p}\+_{1\si} & \vec{p}\+_{2\si} & \dots & \vec{p}\+_{N^{p} \si}
\end{pmatrix}
\end{align}
to get a more compact expression.
The hybridization functions $\Delta^{dd}_{mm'}(i\omega)$, $\Delta^{pp}_{nn'}(i\omega)$ and $\Delta^{dp}_{mm'}(i\omega)$ are related to the parameters in Eq. \eqref{eq:Ham_hyb} by {\color{black}
\begin{align}
\Delta^{dd}_{mm'}(i\omega) &= \sum_{m''k} \frac{V^{dd}_{mm''k}V^{dd*}_{m''m'k}}{i\omega - E^{d}_{m''k}} + \sum_{nk} \frac{V^{dp}_{mnk}V^{pd*}_{nm'k}}{i\omega - E^{p}_{nk}}\\
\Delta^{pp}_{nn'}(i\omega) &= \sum_{n''k} \frac{V^{pp}_{nn''k}V^{pp*}_{n''n'k}}{i\omega - E^{p}_{n''k}} + \sum_{mk}\frac{V^{pd}_{nmk}V^{dp*}_{mn'k}}{i\omega - E^{d}_{mk}}\\
\Delta^{dp}_{mn}(i\omega) &= \sum_{m''k} \frac{V^{dd}_{mm''k}V^{dp*}_{m''nk}}{i\omega - E^{d}_{m''k}} + \sum_{n''k}\frac{V^{dp}_{mn''k}V^{pp*}_{n''nk}}{i\omega - E^{p}_{n''k}}\text{ .}
\end{align}}
They will be determined from the DMFT self-consistency condition, incorporating the structure of the lattice. In this sense, a non-zero $\Delta^{dp}_{mm'}(i\omega)$ arises from inter-site hopping from one shell to another.\par
In the following, it will be convenient to introduce the $(N^{d}+N^{p})\times (N^{d}+N^{p})$ matrix Green's function
\begin{align}
\mathbf{G}_{\si}(\tau) = -\Braket{\begin{pmatrix}
\vec{d}_{\si}(\tau)\\
\vec{p}_{\si}(\tau)
\end{pmatrix}\begin{pmatrix}
\vec{d}\+_{\si}(0) &  \vec{p}\+_{\si}(0)
\end{pmatrix}} \text{ .}
\end{align}

Considering the case of TMO's, it might seem odd to find the ligand p orbitals on the same impurity as the d orbitals corresponding to the transition metal (TM). In realistic materials, such as NiO or the cuprate superconductors, the TM atoms are rather surrounded by their oxygen ligands. The choice of our model can, however, be justified by considering a Zhang-Rice-type construction\cite{ZhangRice}, in which a unitary transformation is applied to change basis from the original ligands to Wannier orbitals, centered on the TMs. {\color{black}For details about the derivation, the reader may refer to the original paper by Zhang and Rice\cite{ZhangRice}; a more pedagogic introduction can be found here\cite{MyThesis}}.\par
Impurity models like \eqref{eq:the_action} constitute complex many-body problems, for which there is, in general, no analytic solution. Still, the marvel of modern computational techniques, like the variety of available quantum Monte Carlo (QMC) solvers, allows for a numerical solution up to (in theory) arbitrary precision. In practice, however, one encounters problems: The computational complexity of most QMC algorithms scale very badly with the number of impurity orbitals (e.g. in the case of the continuous time hybridization expansion solver, the scaling is exponential). To make matters worse, a non-zero intershell hybridization function $\Delta^{dp}_{mm'}(i\omega)$ in \eqref{eq:action_hyb} will give rise to a negative sign problem.\\
The method that we present in this paper is designed to tackle these problems, by considering an efficient hybrid approach, combining QMC with an analytical slave rotor technique. The technique is a direct extension of the shell folding technique, which we shall briefly review in the following.

\subsection{Shell folding}\label{sec:shell_folding}
The shell folding technique is based on the observation that, by means of a purely algebraic manipulation, we can re-write the interacting part of Hamiltonian \eqref{eq:action_int} in a way that lets us separate fluctuations of the total charge from those of the individual orbitals. Indeed, this local Hamiltonian can be written as
\begin{align}\label{eq:action_int_sf}
 \begin{split}
H_{int} &=\frac{1}{2} \sum_{(m\si)\neq(m'\si')}\tilde{U}^{dd}_{m\si m'\si'}\hat{n}^{d}_{m\si}\hat{n}^{d}_{m'\si'}\\
+&\frac{1}{2}\sum_{(n\si)\neq(n'\si')} \tilde{U}^{pp}_{n\si n'\si'}\hat{n}^{p}_{n\si}\hat{n}^{p}_{n'\si'} +  \frac{U^{dp}}{2}(\hat{N}-Q_{0})^{2} \text{ ,} \end{split}
\end{align}
with $\hat{N} = \sum_{\si} (\hat{N}^{d}_{\si} + \hat{N}^{p}_{\si})$ and $\tilde{U}^{(dd/pp)}_{m\si m'\si'} = U^{(dd/pp)}_{m\si m'\si'} - U^{dp}$. $Q_{0}$ is an arbitrary, integer value that we introduced by exploiting our freedom to redefine the chemical potential $\mu$.\par
In this notation, it becomes apparent that the intershell interaction $U^{dp}$ renormalizes the intra-shell interactions, while suppressing fluctuations of the total charge. The shell folding approximation assumes that charge fluctuations of the d-shell are perfectly compensated by those of the p-shell, which implies a constant total charge, such that we can drop the last term in \eqref{eq:action_int_sf}. This way, one arrives at a Hamiltonian in which the d- and p-shell are decoupled. If, furthermore, the p orbitals are effectively treated by a static (Hartree/Hartree-Fock) approximation, the problem is simplified to one where only the d orbitals are interacting.\par
In the context of TMO's, one argument to justify such an approximation is related to the manipulations applied to the lattice model, necessary to derive a single-site impurity model for both the TM and the ligands. Applying a Zhang-Rice-type transformation to obtain TM-centered ligand-type Wannier orbitals, one eventually arrives at a Hamiltonian, in which the amplitude of the inter-site hoppings is rather small compared to the on-site hybridization between the TM and the Wannier orbitals on the same site. This implies that the dominant screening process for TM d orbitals on a given site is provided by charge fluctuations of the ligand-type Wannier orbitals on the same site, 
such that the overall charge of the site remains approximately constant.   

\section{Slave rotor formulation of the d-p model}\label{sec:slave_rotor_derivation}
In the following sections, we present an efficient way to extend the shell folding approximation, by deriving a theory that incorporates fluctuations of the total local charge $N$ in an analytic way. \par
We achieve this by employing the slave rotor technique\cite{SR_impurity,SR_meanfield}, which was invented as a lightweight method to treat strongly correlated systems. The main idea is to re-write the original electronic operators by means of a rotor phase variable $\theta(\tau) \in [0;2\pi)$, carrying the charge, and auxiliary ``pseudo'' fermions, carrying the spin. If the fluctuations of charge and spin live on different energy scales, it is possible to effectively decouple these degrees of freedom in a mean field fashion, which drastically simplifies the problem. \par
In case of our d-p problem, we want an effective way to treat the total charge  corresponding to the operator $\sum_{\si}(\hat{N}_{\si}^{d} + \hat{N}_{\si}^{p})$. Following the logic of the shell-folding approximation, we assume that fluctuations of the total charge live on a different energy scale than the intra-shell fluctuations, therefore motivating us to apply a slave rotor decoupling. 
\subsection{Rotorization}
We start by re-writing the electronic creation/annihilation operators as a product of rotor and pseudo fermionic operators, denoted with a ``tilde''
\begin{align}\label{eq:SR_operatortrans_d}
d\+_{\si} &= \tilde{d}\+_{\si}  e^{i\theta} \quad \text{,}\quad  d_{\si}  = \tilde{d}_{\si}  e^{-i\theta}\\
p\+_{\si}  &= \tilde{p}\+_{\si}  e^{i\theta} \quad \text{,}\quad  p_{\si}  = \tilde{p}_{\si}  e^{-i\theta} \text{ .}\label{eq:SR_operatortrans_p}
\end{align}
The new, composite operators act on an enlarged Hilbert space 
\begin{align}\label{eq:integer_Q0}
\begin{split}
\Ket{\si_{1}^{d}\dots\si^{d}_{Q^{d}},\si^{p}_{1}\dots\si^{p}_{Q^{p}}} = \Ket{\tilde{\si}_{1}^{d}\dots\tilde{\si}^{d}_{Q^{d}},\tilde{\si}^{p}_{1}\dots\tilde{\si}^{p}_{Q^{p}}}\\
\otimes\Ket{ Q^{d} + Q^{p} - Q_{0}}_{\theta}\text{ ,}
\end{split}
\end{align}
where the integer $Q_{0}$ allows to shift the rotor ground state.
For the most simple case of a model with only one d and one p orbital we can for instance set $Q_{0}=2$, such that the rotor ground state corresponds to the half-filled system. We then have states like $\Ket{0,0} = \Ket{0,0}\Ket{-2}_{\theta}$, $\Ket{0,\doa^{p}} = \Ket{0,\tilde{\doa}^{p}}\Ket{-1}_{\theta}$, $\Ket{\upa^{d}, 0} = \Ket{\tilde{\upa}^{d}, 0}\Ket{-1}_{\theta}$, $\Ket{\upa^{d}, \doa^{p}} = \Ket{\tilde{\upa}^{d}, \tilde{\doa}^{p}}\Ket{0}_{\theta}$, $\Ket{\upa^{d}\doa^{d}, \doa^{p}} = \Ket{\tilde{\upa}^{d}\tilde{\doa}^{d}, \tilde{\doa}^{p}}\Ket{1}_{\theta}$, et cetera. \par
The symbols $e^{i\theta}$/$e^{-i\theta}$ can be understood as ladder operators, raising/lowering the angular momentum $l$, which, for physical states, corresponds to the total charge $l\; \widehat{=}\; Q^{d} + Q^{p}-Q_{0}$
\begin{align}\label{eq:apply_rotor_creation}
e^{i\theta}\Ket{l}_{\theta} = \Ket{l + 1}_{\theta} \quad\text{,}\quad e^{-i\theta}\Ket{l}_{\theta} = \Ket{l- 1}_{\theta} \text{ .}
\end{align}
The angular momentum states $\Ket{l}_{\theta}$ are eigenstates of the angular momentum operator $\hat{L} = -i \partial_{\theta}$
\begin{align}
\hat{L} \Ket{l}_{\theta} = l \Ket{l}_{\theta}\text{ ,}
\end{align}
which is conjugate to the phase $[\theta,\hat{L} ] = i$.\par
The purpose of this new formalism is, that it allows us to replace the original total charge operator by this angular momentum operator, acting only on the auxiliary phase variable
\begin{align}\label{eq:L_substitution}
\hat{N} = \sum_{\si}(\hat{N}_{\si}^{d} + \hat{N}_{\si}^{p}) \rightarrow \hat{L}+ Q_{0} \text{ .}
\end{align}
The expanded Hilbert space will also contain states, where the angular momentum does not correspond to the cumulative electronic charge, e.g. $\Ket{0,0} = \Ket{0,0}\Ket{-1}_{\theta}$, $\Ket{0,\doa^{p}} = \Ket{0,\tilde{\doa}^{p}}\Ket{-2}_{\theta}$ or $\Ket{\upa^{d}, 0} = \Ket{\tilde{\upa}^{d}, 0}\Ket{1}_{\theta}$. In order for the substitution \eqref{eq:L_substitution} to be valid, we need to eliminate such ``unphysical'' states. 
This can be done by means of  a projector $\hat{P}_{phys}$, that can be formulated in terms of a static Lagrange multiplier $\phi_{0}$
\begin{align}
\hat{P}_{phys} = \int\mathrm{d}\phi_{0}\; e^{i\phi_{0}(\hat{L} - \hat{N}+Q_{0})} \text{ .}
\end{align} 
This corresponds to adding an additional term to the Hamiltonian 
\begin{align}\label{eq:Ham_constraint}
H \rightarrow H + h(\hat{L} - \hat{N} + Q_{0}) \text{ ,}
\end{align}
where we substituted $h= -i\phi_{0}$. Integrating over $h$, one would exactly cancel all unphysical states and retrieve the physics of our original problem. In practice, however, one rather treats the constraint on average by fixing the Lagrange multiplier to its complex saddle point value. The value of $h$ is then defined by the saddle point condition 
\begin{align}\label{eq:saddle_point_condition}
\langle\hat{L}\rangle_{h} = \langle\hat{N}\rangle_{h} - Q_{0} \text{ .}
\end{align}

Coming back to our d-p problem, the slave rotor formalism gives us the possibility to replace the last term in \eqref{eq:action_int_sf}, which is quartic in the electronic fields (and otherwise dropped within the shell folding approximation), by a simple kinetic term 
\begin{align}
\frac{U^{dp}}{2}(\hat{N}-Q_{0})^{2} \rightarrow \frac{U^{dp}}{2} \hat{L}^{2} \text{ .}
\end{align}

\subsection{Deriving the Slave Rotor Hamiltonian}
We proceed by applying the substitutions \eqref{eq:SR_operatortrans_d}, \eqref{eq:SR_operatortrans_p} and \eqref{eq:L_substitution} to the various parts of the Hamiltonian \eqref{eq:the_action}. A superscript ``$^{SR}$'' will designate the parts of the transformed Hamiltonian. \par
The term incorporating the Lagrange multiplier (Eq. \eqref{eq:Ham_constraint}) is added to the free, atomic part
\begin{align}\label{eq:action_0_SR}
\begin{split}
H_{0}^{SR}&=\sum_{m\si} \tilde{d}\+_{m\si}(-\mu+\eps_{m}^{d} - h)\tilde{d}_{m\si} \\
 &+ \sum_{n\si} \tilde{p}\+_{n\si}(-\mu+\eps_{n}^{p} - h)\tilde{p}_{n\si} \\
 & + \sum_{m n\si} \left( V_{mn}\tilde{d}\+_{m\si}\tilde{p}_{n\si} + h.c.\right)  + h \hat{L}  \text{ ,}
 \end{split}
 \end{align}
which  otherwise does not change, since the rotor operators cancel each other out (e.g. $\tilde{d}\+_{m\si}e^{i\theta}\tilde{p}_{n\si}e^{-i\theta} = \tilde{d}\+_{m\si}\tilde{p}_{n\si}$). \par
This is not the case for the hybridization part Eq. \eqref{eq:Ham_hyb}, which transforms to {\color{black}
\begin{align}\label{eq:Ham_hyb_SR}
\begin{split}
H_{hyb}^{SR} &=   \sum_{mm'k\si} \dt\+_{m\si} V^{dd}_{mm'k} b^{d}_{m'k\si}\rot + h.c\\
&+ \sum_{m n k\si} \left\{\dt\+_{m\si}V^{dp}_{mn k} b^{p}_{nk\si}\rot + \pt\+_{n\si}V^{pd}_{nm k} b^{d}_{mk\si}\rot \right\}  +h.c.\\
&+ \sum_{nn' k\si} \pt\+_{n\si}V^{pp}_{nn'k}b^{p}_{n'k\si}\rot + h.c.\\
&+ \sum_{mk\si} E^{d}_{mk}b^{d\dagger}_{mk\si}b^{d}_{mk\si} + \sum_{nk\si}E^{p}_{nk}b^{p\dagger}_{nk\si} b^{p}_{nk\si}\text{ ,}
\end{split}
\end{align}}

where the time dependence of the phase variable $\theta(\tau)$ causes 
an explicit coupling between the pseudo fermions and rotor variables. 
Finally, we have got the interaction part
\begin{align}\label{eq:action_int_SR}
 \begin{split}
H^{SR}_{int} &= \frac{1}{2} \sum_{(m\si)\neq(m'\si')}\tilde{U}^{dd}_{m\si m'\si'}\hat{n}^{\dt}_{m\si}\hat{n}^{\dt}_{m'\si'}  \\
+&\frac{1}{2}\sum_{(n\si)\neq(n'\si')} \tilde{U}^{pp}_{n\si n'\si'}\hat{n}^{\pt}_{n\si}\hat{n}^{\pt}_{n'\si'} +  \frac{U^{dp}}{2}\hat{L}^{2} \text{ ,} \end{split}
\end{align}
where we re-expressed the last term using the slave rotor angular momentum variable. 
\subsubsection{Mean-field decoupling}\label{sec:mean_field_eq}
Until now, the only approximation that we employed was to fix the Lagrange multiplier to its saddle point value $h$; otherwise everything is still treated exactly. Re-writing the Hamiltonian \eqref{eq:the_action} in terms of the slave rotor formalism brought us one step closer to an effective, dynamic description of total charge fluctuations beyond shell folding. However, we traded a quartic fermionic term in \eqref{eq:action_int_SR} with an explicit interaction between slave rotor and pseudo fermionic variables in \eqref{eq:Ham_hyb_SR}. \par
To make the problem tractable, we follow the strategy presented in Ref. \onlinecite{SR_meanfield} and decouple $H^{SR}$ in a mean-field fashion
\begin{align}\label{eq:SR_meanfield}
H^{SR}_{\text{mean-field}} = H^{f} + H^{\theta}\text{ .}
\end{align}
The pseudo fermionic Hamiltonian $H^{f}$ reads
\begin{align}\label{eq:Ham_hyb_f}
H^{f} = H^{f}_{0} + H^{f}_{int}+ H^{f}_{hyb}\text{ ,}
\end{align}
with
\begin{align}
\begin{split}\label{eq:Hf_H0}
\color{black}H^{f}_{0} &= \color{black}  \sum_{m\si} \tilde{d}\+_{m\si}(-\mu+\eps_{m}^{d} - h)\tilde{d}_{m\si}\\ 
& \color{black}+ \sum_{n\si} \tilde{p}\+_{n\si}(-\mu+\eps_{n}^{p} - h)\tilde{p}_{n\si}\\
&\color{black}+ \sum_{m n\si} \left( V_{mn}\tilde{d}\+_{m\si}\tilde{p}_{n\si} + h.c.\right)\text{ ,}
\end{split}\\
\begin{split}\label{eq:Hf_Hint}
H^{f}_{int}&=\frac{1}{2} \sum_{(m\si)\neq(m'\si')}\tilde{U}^{dd}_{m\si m'\si'}\hat{n}^{\dt}_{m\si}\hat{n}^{\dt}_{m'\si'}\\
&+\frac{1}{2}\sum_{(n\si)\neq(n'\si')} \tilde{U}^{pp}_{n\si n'\si'}\hat{n}^{\pt}_{n\si}\hat{n}^{\pt}_{n'\si'}
\end{split}\\
\begin{split}\label{eq:Hf_Hhyb}
\color{black}H^{f}_{hyb}&=\color{black}\sum_{mm'k\si} \dt\+_{m\si} V^{dd}_{mm'k} b^{d}_{m'k\si}\langle\rot\rangle_{\theta}+ h.c\\
&\color{black}+ \sum_{m n k\si} \left\{\dt\+_{m\si}V^{dp}_{mn k} b^{p}_{nk\si} + \pt\+_{n\si}V^{pd}_{nm k} b^{d}_{mk\si} \right\}\langle\rot\rangle_{\theta}  +h.c.\\
&\color{black}+ \sum_{nn' k\si} \pt\+_{n\si}V^{pp}_{nn'k}b^{p}_{n'k\si}\langle\rot\rangle_{\theta} + h.c.\\
&\color{black}+ \sum_{mk\si} E^{d}_{mk}b^{d\dagger}_{mk\si}b^{d}_{mk\si} + \sum_{nk\si}E^{p}_{nk}b^{p\dagger}_{nk\si} b^{p}_{nk\si}\text{ .}
\end{split}
\end{align}
The decisive difference of this Hamiltonian compared to the original one \eqref{eq:the_action} is, that the new expressions do not contain any intershell interactions $\sim U^{dp}$. Apart from that, other differences include:
\begin{enumerate}
\item Compared to \eqref{eq:action_0}, the bare energies of the d and p orbitals are shifted by the Lagrange multiplier $h$, which emerged from the saddle point condition \eqref{eq:saddle_point_condition}.
\item Since we adopted the shell folding notation, the interaction strengths in \eqref{eq:Hf_Hint} are reduced by the the intershell interaction 
\begin{align}
\tilde{U}^{dd} &=U^{dd} - U^{dp}\\
\tilde{U}^{pp} &=U^{pp} - U^{dp} \text{ .}
\end{align} \par
This can be further exploited for an effective treatment of the p-p interactions. 
\item The hybridization amplitudes with the bath sites are renormalized by $\langle\rot\rangle$.
\end{enumerate}  \par
On the other hand, the rotor part of the mean-field Hamiltonian \eqref{eq:Ham_hyb_f} takes the simple form
\begin{align}\label{eq:H_theta}
H^{\theta} = h\hat{L} + \frac{U^{dp}}{2}\hat{L}^{2} + \frac{1}{2}E_{kin}^{f}\left(\rot + \mrot\right) \text{ ,}
\end{align}
where $E_{kin}^{f}$ is the kinetic energy of the pseudo fermionic system, defined as
\begin{align}\label{eq:Ekin_def}
\begin{split}
\frac{E_{kin}^{f}}{2} &= \sum_{mm'k\si} \Braket{\dt\+_{m\si} V^{dd}_{mm'k} b^{d}_{m'k\si}}_{f}\\
&+ \sum_{m n k\si} \braket{\dt\+_{m\si}V^{dp}_{mn k} b^{p}_{nk\si}}_{f} + \braket{\pt\+_{n\si}V^{pd}_{nm k} b^{d}_{mk\si}}_{f}\\
&+ \sum_{nn' k\si} \braket{\pt\+_{n\si}V^{pp}_{nn'k}b^{p}_{n'k\si}}_{f}\\
&= \frac{1}{\beta}\sum_{i\omega\si} e^{i\omega0^{+}}\Tr\{ \mathbf{G}^{f}_{\si}(i\omega)\mathbf{\Delta}(i\omega) \}\text{ .}
\end{split}
\end{align}
The label ``f'' indicates that these quantities correspond to the pseudo fermionic part of the mean-field Hamiltonian \eqref{eq:SR_meanfield}.\par
Within the mean-field formalism, the physical Green's function factorizes to 
\begin{align}\label{eq:G_factorize_mean_field}
\mathbf{G}_{\si}(\tau) = \mathbf{G}_{\si}^{f}(\tau)G^{\theta}(\tau) \text{ ,}
\end{align}
where the pseudo fermion and the rotor Green's function 
\begin{align}
\mathbf{G}^{f}_{\si}(\tau) &= -\Braket{\begin{pmatrix}
\vec{\dt}_{\si}(\tau)\\
\vec{\pt}_{\si}(\tau)
\end{pmatrix}\begin{pmatrix}
\vec{\dt}\+_{\si}(0) &  \vec{\pt}\+_{\si}(0)
\end{pmatrix}}_{f} \text{ ,}\\
G^{\theta}(\tau) &= \Braket{e^{-i(\theta(\tau) - \theta(0))}}_{\theta} \text{ ,}\label{eq:Rotor_Greensfun}
\end{align}
are defined with respect to the corresponding parts $H^{f}$ and $H^{\theta}$ of the mean-field Hamiltonian \eqref{eq:SR_meanfield}.\par
\begin{figure}
\includegraphics[width=0.48\textwidth]{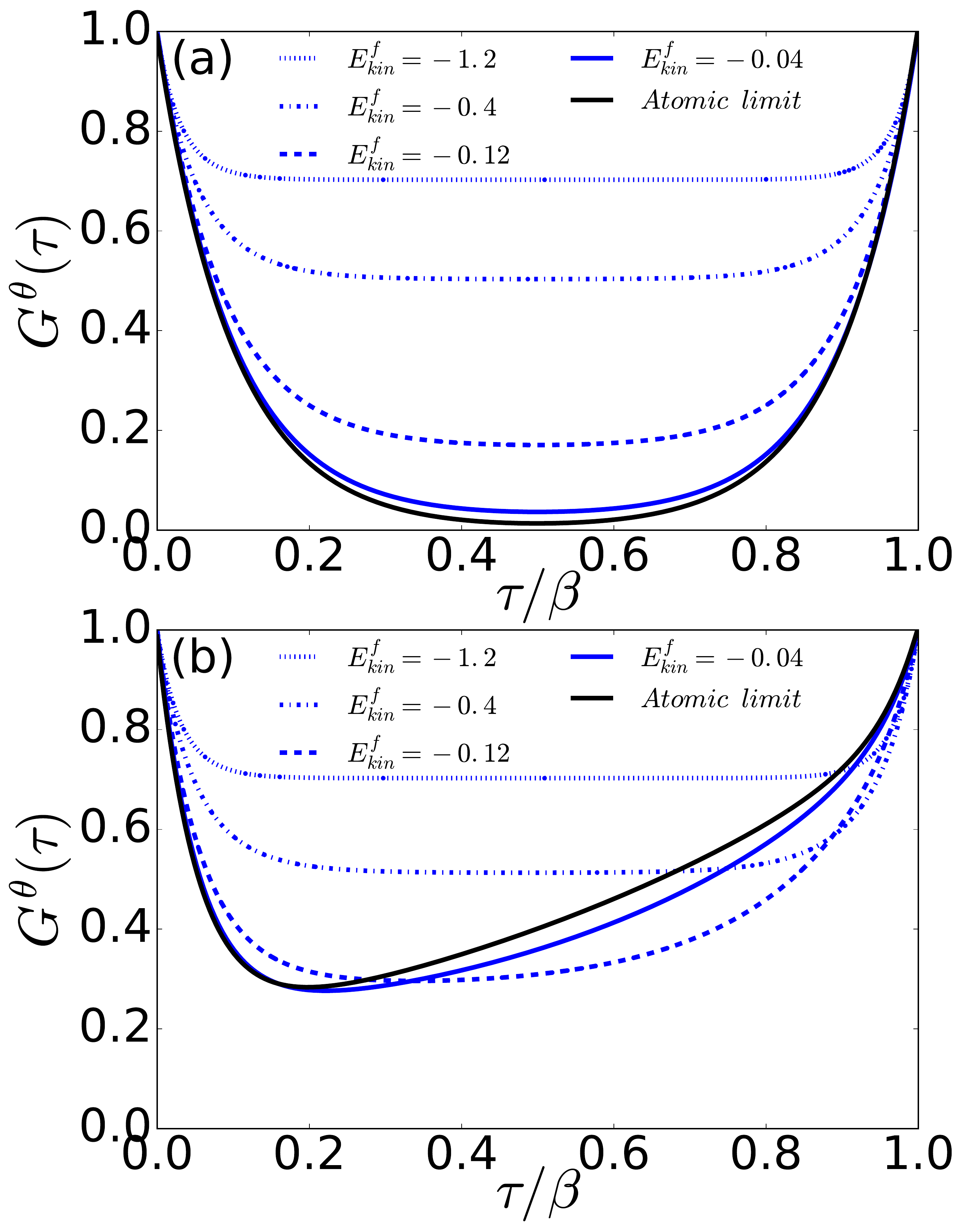}
\caption{Rotor Green's functions $G^{\theta}(\tau)$ for various parameters $E^{f}_{kin}$ (the ``Atomic limit'' is equal to $E^{f}_{kin}=0$), $U^{dp}=0.5$, $\beta=40$ and $Q_{0}=2$. Panel (a) was evaluated at half filling $\langle\hat{L}\rangle_{H_{\theta}} = 0$, while panel (b) corresponds to the incommensurate filling  $\langle\hat{L}\rangle_{H_{\theta}} = 0.2$. Note how the asymmetry caused by finite dopings is ``washed out'' by the last term in \eqref{eq:H_theta} which is proportional to $E^{f}_{kin}$.}\label{fig:Gtheta_tau}
\end{figure}
For a given set of hybridization strengths $V$'s and bath-site energies, 
the mean-field condition requires a self-consistent evaluation of the quantities $\langle\rot\rangle_{\theta}$ and $E_{kin}^{f}$, while $h$ must be set such that saddle point condition \eqref{eq:saddle_point_condition} is fulfilled.\par
Note that 
evaluating $E^{f}_{kin}$ still requires calculating $\mathbf{G}^{f}_{\si}(\tau)$ from the interacting impurity model \eqref{eq:Ham_hyb_f}. As compared to the original problem \eqref{eq:the_action}, the complexity of such calculations can be drastically reduced by acknowledging that often\cite{shell_folding,Hansmann_2014}, the intershell interaction $U^{dp}$ is of a similar magnitude as $U^{pp}$, such that $\tilde{U}^{pp} = U^{pp}-U^{dp}$ is small. The weak remaining interaction can be efficiently treated within a Hartree approximation. In this way, we end up with an quantum impurity model, in which the p orbitals are non-interacting and can, therefore, be integrated out. \par
On the other hand, the rotor Green's function \eqref{eq:Rotor_Greensfun} can be evaluated with minimal computational cost by diagonalizing the rotor Hamiltonian \eqref{eq:H_theta}. This requires a truncation of the infinite spectrum of the rotor operator $\hat{L}$, which can then be written as a matrix of finite dimension. Upon increasing the dimension of the matrix, the results quickly converge towards those of the full $\hat{L}$. FIG. \ref{fig:Gtheta_tau} shows the rotor Green's functions for different values of $E^{f}_{kin}$ and different fillings. 
\subsubsection{Spectral properties}
In connection to realistic material investigations, the spectral function $A(\omega) = -\frac{1}{\pi}\mathrm{Im}G(\omega)$ is of central interest, since it can be directly related to experimental photoemission and absorption spectra. The spectral function can be interpreted as a probability distribution function, its integral thus has to be normalized to 1. This normalization is related to the anti-commutation relation of the electron operators $\{d, d\+\} = 1$, $\{p, p\+\} = 1$. By construction, this physical constraint is also fulfilled within the slave rotor formalism, since $\{\dt e^{-i\theta}, \dt\+e^{i\theta}\} = \{\dt, \dt\+\} =1$ and $\{\pt e^{-i\theta}, \pt\+e^{i\theta}\} = \{\pt, \pt\+\} =1$.
This still holds true for the mean-field implementation pursued in this paper. Indeed, using the Lehmann representation of the rotor Green's function in the real frequency domain $G^{\theta}(\omega)$, one can easily show that the mean-field-induced factorization \eqref{eq:G_factorize_mean_field} leads to a redistribution of spectral weight corresponding to the pseudo fermionic Green's function $\mathbf{G}^{f}_{\si}(\omega)$.

{\color{black}
\subsubsection{Symmetry of the Rotor Green's function}
It is interesting to note that the rotor Green's function is invariant under changes of the parameter $h$ that change the angular momentum $l$ by an integer value. 
This is most easily seen by considering the following transformation on the rotor Hamiltonian \eqref{eq:H_theta}
\begin{align}
\begin{split}
&H^{\theta} \rightarrow \tilde{H}^{\theta} = e^{-i\theta}H^{\theta}e^{i\theta}\\
&=h(\hat{L}+1) + \frac{U^{dp}}{2}(\hat{L}+1)^{2} + \frac{1}{2}E_{kin}^{f}\left(\rot + \mrot\right)  \text{ .}
\end{split}
\end{align}
Applying this transformation is equivalent to changing $h \rightarrow h + U^{dp}$, and has the effect of decreasing the expectation value of the angular momentum
\begin{align}
\langle \hat{L}\rangle_{\tilde{H^{\theta}}} = \frac{1}{Z}\Tr\{ e^{-i\theta}e^{-\beta H}e^{i\theta}\hat{L}\} = \langle\hat{L}\rangle_{H^{\theta}} - 1\text{ .}
\end{align}
On the other hand, it is straightforward to show that applying this transformation to the Hamiltonian leaves the Green's function unchanged
\begin{align}\label{eq:rotor_symmery}
\begin{split}
\Braket{e^{-i(\theta(\tau) - \theta(0))}}_{\tilde{H}^{\theta}} = &\Braket{e^{-i(\theta(\tau) - \theta(0))}}_{H^{\theta}} \text{ .}
\end{split}
\end{align}
This means, that the rotor Green's function is invariant under $h\rightarrow h+nU^{dp}$, where $n\in \mathbb{Z}$ is an arbitrary integer. Furthermore, it implies that the symmetry property $G^{\theta}(\beta/2 + \tau) = G^{\theta}(\beta/2 - \tau)$, that can be observed in FIG. \ref{fig:Gtheta_tau} (a) for a half filled system with $h=0$ is periodically restored for $h=nU^{dp}$, $n\in\mathbb{Z}$.

\subsubsection{Simplifications at commensurate fillings}\label{sec:commensurate_simple}
The above observations have some very practical implications. In particular, they allow for simplifications in the calculation of the rotor Green's function at arbitrary integer fillings, which would otherwise only occur at half-filled systems due to particle-hole symmetry.\\
In the case of a half-filled system, the constraint \eqref{eq:saddle_point_condition} can be trivially fulfilled by setting $h=0$, if the integer $Q_{0}$ is set equal to the number of orbitals. This holds true, irrespective of the value of $E^{f}_{kin}$.\\
Property \eqref{eq:rotor_symmery} described in the previous section now implies, that the rotor Green's function, at any commensurate filling $\langle \hat{N}\rangle$, is identical to the one at half filling, corresponding to Hamiltonian \eqref{eq:H_theta} with $h=0$.\\
Another way of seeing this is by reconsidering our freedom to chose the integer $Q_{0}$ in \eqref{eq:integer_Q0}.
In the case of commensurate fillings $\langle \hat{N}\rangle$, we can always set $Q_{0} = \langle \hat{N}\rangle $, such that the saddle point condition \eqref{eq:saddle_point_condition} is trivially fulfilled by a vanishing Lagrange multiplier $h=0$, since
\begin{align}
\langle\hat{L}\rangle_{h=0} = \langle \hat{N}\rangle - Q_{0} = 0 \text{ .}
\end{align}
In order to perform calculations at arbitrary commensurate fillings, it is thus sufficient to change the definition of the integer $Q_{0}$ and set $h=0$.}

\subsection{Slave Rotor + Dynamical Mean Field Theory}\label{sec:SR_plus_DMFT}
\begin{figure}
\includegraphics[width=0.48\textwidth]{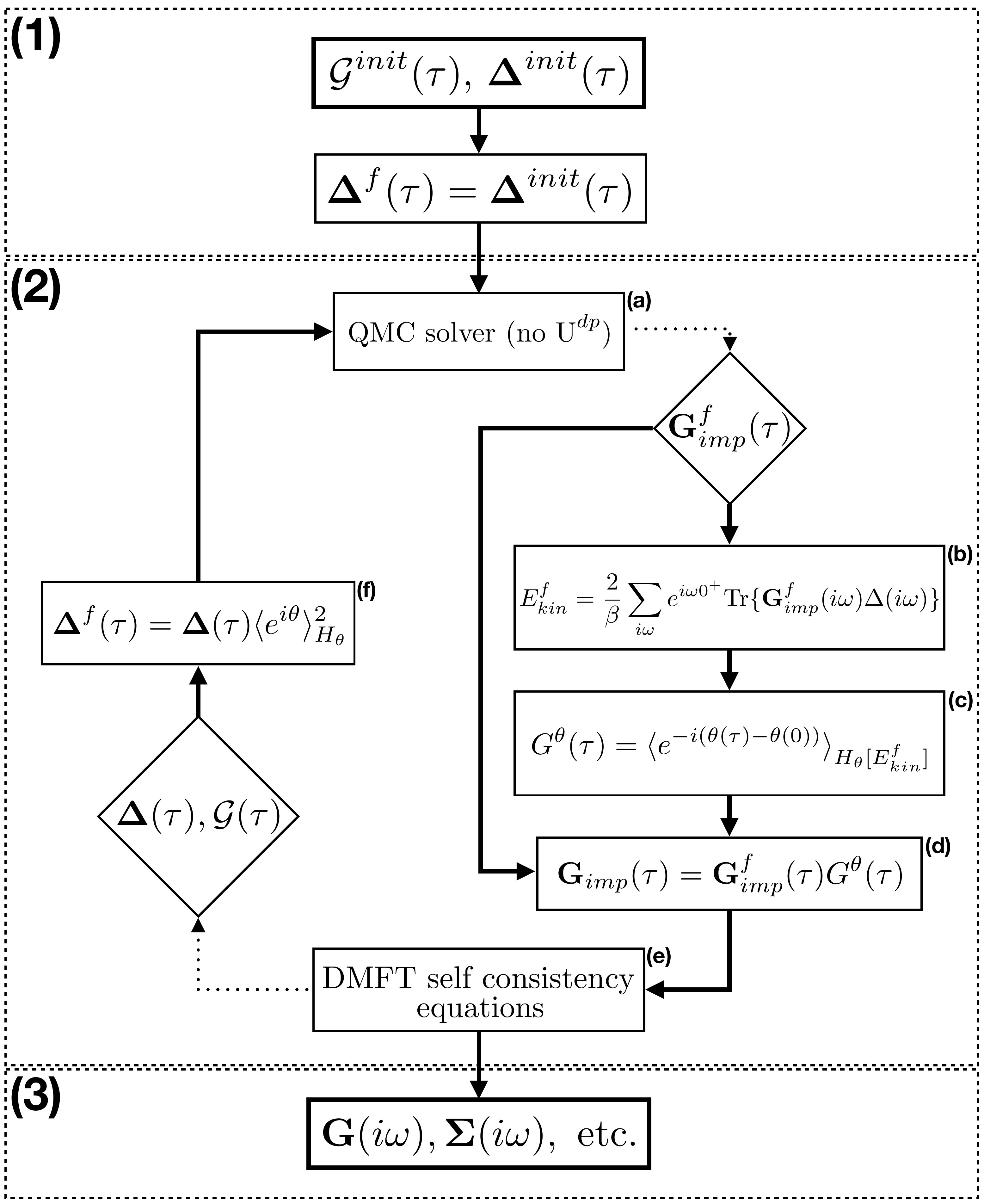}
\caption{\color{black}Scheme of the DMFT loop with the slave rotor approximation. Thick boxes enframe initial inputs and overall results, while calculation steps are put in thin boxes. Diamonds enframe results of calculations from the preceding boxes (to which they are connected by dotted lines) if those calculations are too lengthy to be written out explicitely. For the sake of compactness, spin and orbital indices are omitted. Details on the calculations can be found in the text.}\label{fig:DMFT_cycles}
\end{figure}
In its essence, the slave rotor method described above can be seen as an efficient, approximate quantum impurity solver for the d-p problem described by Hamiltonian \eqref{eq:the_action}. While the physics of impurity models is interesting by itself, much richer phenomena can be expected to emerge in lattice systems. \par
To describe such systems, our slave rotor method can be implemented in the framework of dynamical mean field theory\cite{DMFTkotliar_georges} (DMFT). \par
\begin{figure*}[ht!]
\includegraphics[width=0.8\textwidth]{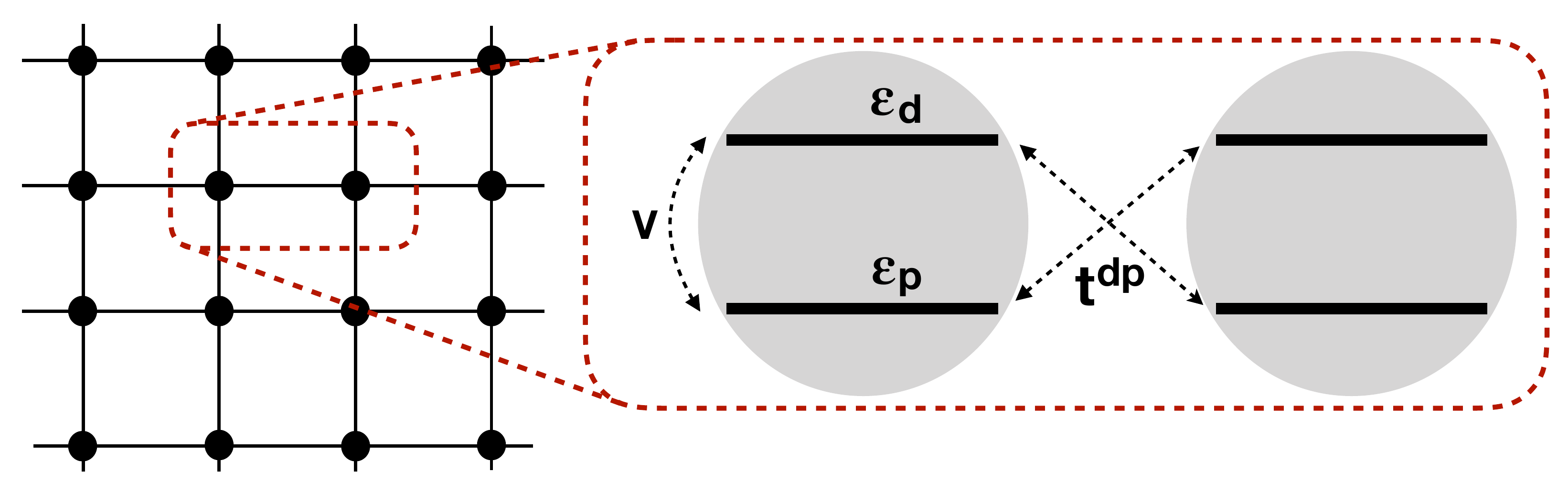}
\caption{Schematic representation of the two-orbital d-p model on a two dimensional square lattice. Up to the chemical potential $\mu$, the d- and p-orbitals on each site are characterized by bare energy levels $\epsilon^{d}$ and $\epsilon^{p}$, respectively, with an on-site hybridization of strength $V$. In the spirit of the Zhang-Rice model\cite{ZhangRice}, nearest-neighbor hoppings are d-to-p only, with a hopping-strength $t^{dp}$.}\label{fig:Udp_model}
\end{figure*}

This can be done in various ways. The most direct procedure would be to solve the mean-field system (Sec. \ref{sec:mean_field_eq}) at every cycle of the DMFT loop. However, such an approach is conceivably inefficient, since solving the mean-field equations involves several runs of quantum Monte Carlo to get $\mathbf{G}^{f}$ from Hamiltonian \eqref{eq:Ham_hyb_f}.\par
A more efficient strategy is to solve the mean-field equations at the same time as the DMFT self-consistency equations, by updating $\langle\rot\rangle_{\theta}$, $E_{kin}^{f}$ and $h$ within the DMFT cycles. \par
The procedure is illustrated in FIG. \ref{fig:DMFT_cycles}, and consists of the following steps:
\begin{table}
\begin{enumerate}
\item \textbf{Initialization:} The DMFT calculation is started with some initial guess for the  Weiss field $\mathbf{\mathcal{G}}^{init}$ or the hybridization function $\mathbf{\Delta}^{init}$ for the physical system. The pseudo fermionic hybridization function $\mathbf{\Delta}^{f}$ is set to $\mathbf{\Delta}^{init}$.
\item \textbf{Self-consistency loop:}
\begin{enumerate}
\item The hybridization function $\mathbf{\Delta}^{f}$ is transferred to the QMC solver, which evaluates the impurity Green's function $\mathbf{G}^{f}$ corresponding to Hamiltonian \eqref{eq:Ham_hyb_f}.
\item From the impurity Green's function, we calculate the kinetic energy of the pseudo fermions via Eq. \eqref{eq:Ekin_def}.
\item Using $E^{f}_{kin}$, the rotor Green's function $G^{\theta}$ is evaluated from Hamiltonian \eqref{eq:H_theta}. FIG. \ref{fig:Gtheta_tau} illustrates $G^{\theta}(\tau)$ for two different fillings and various values of $E^{f}_{kin}$. 
\item The pseudo fermion- and rotor Green's functions are used to reconstruct the physical Green's function $\mathbf{G}(\tau) = \mathbf{G}^{f}(\tau)G^{\theta}(\tau)$, which is used to
\item calculate the self energy $\mathbf{\Sigma}$, the local Green's function $\mathbf{G}_{loc}$ and, finally, the updates for the hybridization function $\mathbf{\Delta}$ and the Weiss field $\mathbf{\mathcal{G}}$.
\item The pseudo fermionic hybrdization function is calculated as $\mathbf{\Delta}^{f}(\tau) = \mathbf{\Delta}(\tau)\langle\rot\rangle^{2}_{H_{\theta}}$, and transferred back to the QMC solver. \par
\end{enumerate}
\item \textbf{Results:} The cycle is converged when the DMFT self-consistency condition $\mathbf{G}_{loc} = \mathbf{G}_{imp}$ is fulfilled, and the values of $\langle\rot\rangle_{H_{\theta}}$ and $E^{f}_{kin}$ are stabilized.
If this is the case, the quantities of interest $\mathbf{G}(i\omega)$, $\mathbf{\Sigma}(i\omega)$, etc. can be extracted. 
\end{enumerate}
We note that FIG. \ref{fig:DMFT_cycles} describes only one possible implementation of the slave rotor + DMFT method, leaving space for optimizations. In practice, e.g., it turned out that convergence can be further stabilized by updating $E^{f}_{kin}$ (step (2) (b)) and thus $G^{\theta}$ only once every 3-5 cycles.
\end{table}

\section{Testing the slave rotor method}\label{sec:Testing}
\subsection{Two orbital d-p model on the square lattice}\label{sec:the_test_model}

In order to test the performance of our slave rotor method, 
we consider a minimal two orbital model. 
These two orbitals serve as proxies for two different atomic shells; as before we shall refer to them as ``d'' and ``p''. The Hamiltonian for our impurity model thus reads
\begin{align}\label{eq:test_model}
H = H_{0} + H_{hyb} + H_{int} \text{ ,}
\end{align}
with
\begin{align}
\begin{split}
H_{0} &= \sum_{\si}(\epsilon^{d} - \mu) d\+_{\si} d_{\si} + \sum_{\si}(\epsilon^{p} - \mu) p\+_{\si} p_{\si}\\
& + \sum_{\si} \left(V d\+_{\si} p_{\si} + h.c. \right) \text{ ,}
\end{split}
\end{align}
\begin{align}
\begin{split}
H_{hyb} &=   \sum_{k\si} \left\{\begin{bmatrix}
d\+_{\si} & p\+_{\si}
\end{bmatrix}  \begin{bmatrix}
V_{k}^{dd}& V_{k}^{dp}\\
V_{k}^{pd}& V_{k}^{pp}
\end{bmatrix} \begin{bmatrix}
b^{d}_{k\si}\\
b^{p}_{k\si}
\end{bmatrix}+ h.c. \right\} \\
&+ \sum_{k\si} E^{d}_{k} b^{d\dagger}_{k\si}b^{d}_{k\si} + \sum_{k\si}E^{p}_{k}b^{p\dagger}_{k\si}b^{p\dagger}_{k\si} 
\end{split}
\end{align}
and 
\begin{align}
\begin{split}
H_{int} &=  (U^{dd}-U^{dp}) n^{d}_{\upa}n^{d}_{\doa} + (U^{pp}-U^{dp})n^{p}_{\upa}n^{p}_{\doa}\\
&+ \frac{U^{dp}}{2}\left(\hat{N}-Q_{0}\right)^{2} \text{ .}
\end{split}
\end{align}
The energies $E^{d/p}_{k}$ of the bath sites, as well as the hybridization matrix $\mathbf{V}_{k}$ are defined via the DMFT self-consistency equations and thus incorporate the structure of the lattice. 
In the following, we consider a two dimensional square lattice with on-site d-p hybridization $V$ and intershell nearest neighbor hopping $t^{dp}$, as depicted in FIG. \ref{fig:Udp_model}. For the sake of simplicity, we will focus on the case where $U^{dp}=U^{pp}$, such that $\tilde{U}^{pp}=U^{pp}-U^{dp} = 0$. The hopping $t^{dp}$ will serve as our unit of energy and will thus be set to $t^{dp}=1$. \par
The form of this model is reminiscent of the one resulting from the Zhang-Rice construction\cite{ZhangRice} for the copper-oxide planes in Cuprate superconductors. In their paper\cite{ZhangRice}, the construction is used to derive an effective two-orbital model on a two-dimensional square lattice, comprised of copper d$_{x^{2}-y^{2}}$ orbitals that hybridize with Wannier orbitals (formed as linear combinations of the oxygen p$_{x}$ and p$_{y}$ orbitals), centered at the d sites. In contrast to our model, the resulting system also includes hopping from the d orbitals to Wannier orbitals beyond nearest-neighbor sites, with an hopping amplitude that decreases as $|t_{ij}| \sim |r_{i}-r_{j}|^{-3}$.\par

\subsection{Implementation and Benchmarking}
\begin{figure}
\includegraphics[width=0.48\textwidth]{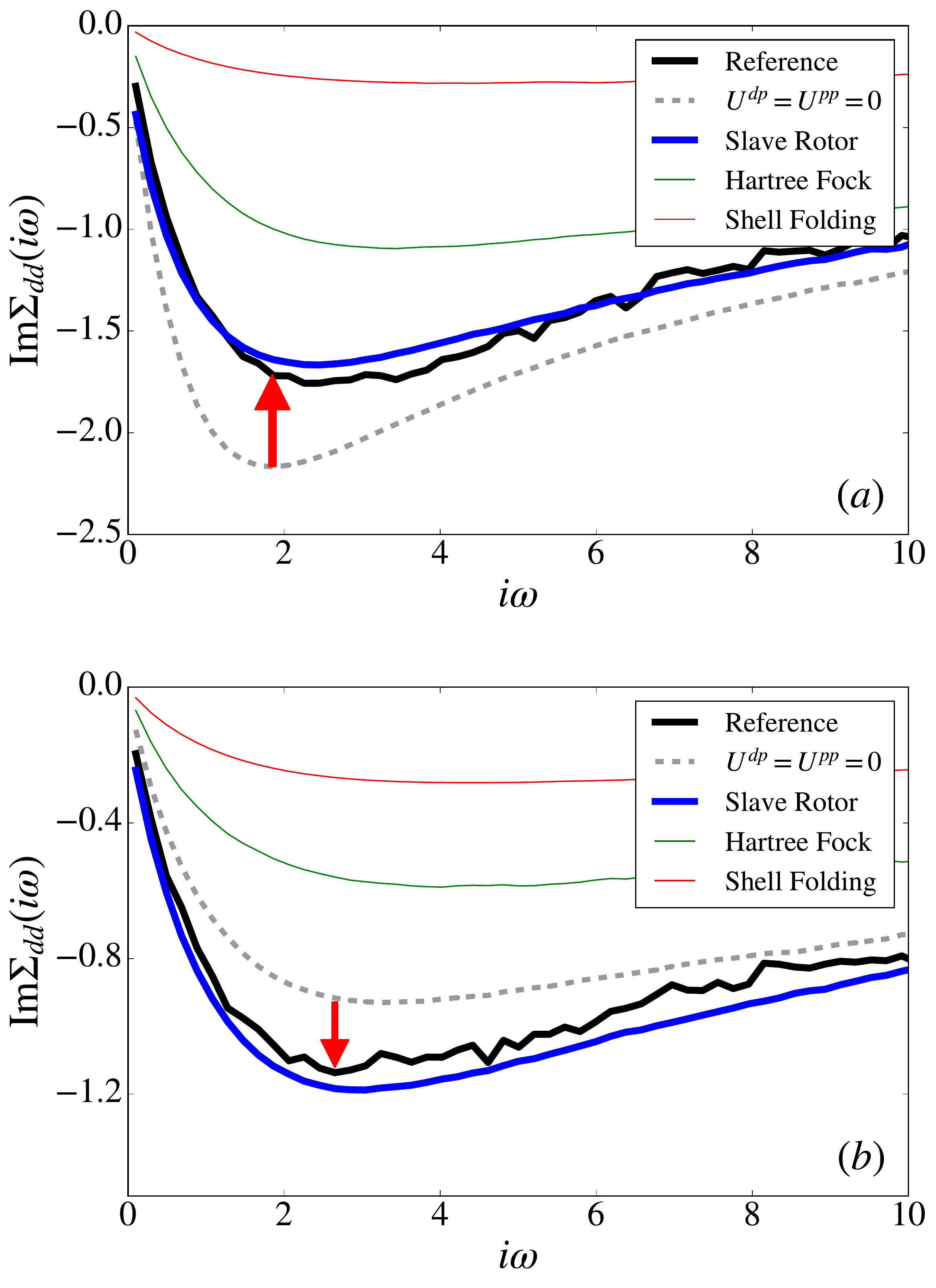}
\caption{Imaginary parts of the d orbital self energy, comparing the QMC results for the full Hamiltonian \eqref{eq:test_model} (reference) to results with $U^{dp}=U^{pp}=0$ and to results obtained from different approximate techniques to treat the intershell interactions. Filling $N=3$, model parameters $U^{dd}=8.75$, $U^{dp}=U^{pp}=0.5U^{dd}$, $\Delta=0$, $\beta=32$. The on-site hybridization is $V=3.5$ panel (a) and $V=7$ in panel (b); the red arrows indicate the direction in which the results are changed by the intershell interactions, corresponding to a decrease (a) or an increase (b) in correlations.}\label{fig:corr_increase_decrease}
\end{figure}

In order to assess the performance of our slave rotor technique, it is benchmarked against reference results for which the full model Hamiltonian \eqref{eq:test_model} -- including intershell interactions -- was solved with quantum Monte Carlo (QMC) techniques. In contrast to the other calculations, the reference curves are often afflicted by relatively high levels of noise. This is due to the negative sign problem, which emerges in QMC calculations for the interacting two-orbital model with off-diagonal hybridization. In order to contain the problem and render the DMFT calculations feasible, we rely on transformations of the orbital basis\cite{QMC_negativesign}.\par
In addition to the reference calculations, the results of the slave rotor technique are compared to those obtained from the model \eqref{eq:test_model} with $U^{dp}=U^{pp}=0$, as well as results from the shell folding scheme described in Section \ref{sec:shell_folding} and from the Hartree-Fock approximation\footnote{Within the Hartree-Fock approximation, the p-p and intershell interactions are approximated on a mean-field level. Neglecting any spin-flipping terms, we get
\begin{align*}\label{eq:Hartree_Fock_dp}
\begin{split}
U^{pp} n^{p}_{\upa}n^{p}_{\doa} &+ U^{dp} (n^{d}_{\upa} + n^{d}_{\doa})(n^{p}_{\upa} + n^{p}_{\doa})\\
& \approx U^{pp} \left\{ n^{p}_{\upa}\Braket{n^{p}_{\doa}} + \Braket{n^{p}_{\upa}}n^{p}_{\doa}  + \Braket{n^{p}_{\upa}}\Braket{n^{p}_{\doa}}  \right\}\\
&+U^{dp} \left\{(n^{d}_{\upa} + n^{d}_{\doa}) \Braket{n^{p}_{\upa} + n^{p}_{\doa}} + \Braket{n^{d}_{\upa} + n^{d}_{\doa}}(n^{p}_{\upa}+ n^{p}_{\doa})\right.\\
&\left. + \Braket{n^{d}_{\upa} + n^{d}_{\doa}}\Braket{n^{p}_{\upa} + n^{p}_{\doa}}\right\}\\
&-U^{dp}\left\{ d\+_{\upa}p_{\upa}\Braket{p\+_{\upa}d_{\upa}} +   \Braket{d\+_{\upa}p_{\upa}}p\+_{\upa}d_{\upa} +  \Braket{d\+_{\upa}p_{\upa}}\Braket{p\+_{\upa}d_{\upa}}\right\}\\
&-U^{dp}\left\{ d\+_{\doa}p_{\doa}\Braket{p\+_{\doa}d_{\doa}} +   \Braket{d\+_{\doa}p_{\doa}}p\+_{\doa}d_{\doa} +  \Braket{d\+_{\doa}p_{\doa}}\Braket{p\+_{\doa}d_{\doa}}\right\} \text{ .}
\end{split}
\end{align*}
Within this approximation, the p electrons are thus treated as being effectively uncorrelated.}. Considering the case where $U^{dp}=U^{pp}$, both of the latter methods reduce the system to one where the p orbitals are effectively uncorrelated. The p electron fields can thus be integrated out and the problem is reduced to correlated one-orbital model, which can be solved with quantum Monte Carlo techniques, without encountering a negative sign problem. This last point also holds true for the implementation of our slave rotor technique; in this case, however, p electron correlations are included via the slave rotors. 

\subsection{Representative results}
In this section, we provide a proof of concept of our novel slave rotor method. For the sake of readability, we will only restrict ourselves to the most important results; a more exhaustive benchmark can be found in Appendix \ref{sec:A1}, where our method is tested in various parameter regimes. \par
Our slave rotor technique is a direct extension of the shell folding approach.  As a first application, we thus show that it is able to cure the most important deficiencies of its progenitor: Being based on the assumption of generalized perfect screening, the shell folding scheme will always predict intershell interactions to decrease the correlations of the system. However, intershell interactions also lead to an increased energy cost for non-local charge excitations, thus increasing the electronic correlations. This can be demonstrated to hold true for finite on-site hybridization $V\neq0$ by considering the atomic limit (see Appendix \ref{sec:atomic_limit_energy} and Ref. \onlinecite{Jakob_thesis}).\par
\begin{figure*}
\includegraphics[width=\textwidth]{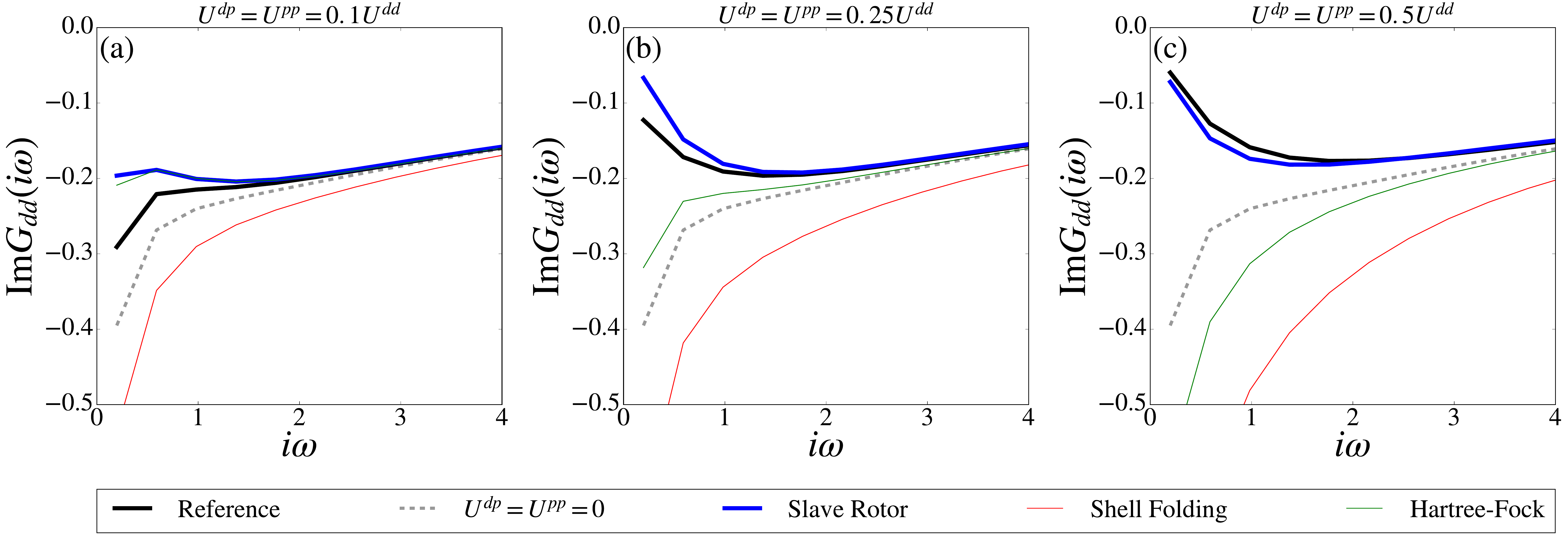}
\caption{Imaginary part of the d orbital Green's function for different values of the intershell and p-p interaction, comparing the QMC results for the full Hamiltonian \eqref{eq:test_model} to different approximate methods. Panels (a)-(c) show the results for increasing $U^{dp}$ and $U^{pp}$; the total filling is $N=3$, individual fillings are fixed to  $\sum_{\si} \langle n^{d}_{\si}\rangle = 1.1$ and $\sum_{\si} \langle n^{p}_{\si}\rangle = 1.9$. The residual model parameters are $U^{dd}=6.5$, $V=3.5$, $\beta=16$ ($\mu$ and $\Delta$ are fixed corresponding to the fillings).}\label{fig:MIT}
\end{figure*}
First, we shall consider the case $\langle\hat{N}\rangle=3$, which is motivated by the physics of transition metal oxides with partially filled d-bands and filled p-band (e.g. undoped CuO$_{2}$ planes\cite{ZhangRice,MyThesis}). In this case, we can make use of the simplifications that come along in the case of commensurate fillings (see Sec. \ref{sec:commensurate_simple}).  
\subsubsection{Two faces of intershell interactions}
FIG. \ref{fig:corr_increase_decrease} illustrates the competition between these two effects by comparing the change in the imaginary parts of the d orbital self energies upon turning on intershell interactions for two different parameter sets. In panel (a), the on-site hybridization is set to $V=3.5$, and intershell interactions decrease the correlations (red arrow up). On the other hand, setting $V=7$, the effect is reversed and the additional interactions lead to an increase in correlations (red arrow down). \par
In both cases, the curve corresponding to the slave rotor approximation is rather close to the reference solution, correctly reproducing the effect of $U^{dp}$ and $U^{pp}$ for the different parameters. In contrast, the shell folding scheme drastically overestimates the decrease in correlations for both parameter sets. This is also the case for the Hartree-Fock approximations. While the corresponding results are better than the ones obtained from the shell folding approach, their deviations from the reference solutions are still bigger than those from the curves obtained from setting $U^{dp}=U^{pp}=0$. This indicates error cancellation among higher-order terms and underlines the fact that naive low-order corrections can be worse than simply ignoring the intershell interactions.\par
\subsubsection{Increasing the intershell interactions: The metal insulator transition}
In FIG. \ref{fig:MIT} we show results from different methods for different strengths of $U^{dp}$ ($=U^{pp}$). One effect of the intershell interactions is the redistribution of charge, induced by the effective change of the charge transfer energy $\Delta$ due to the corresponding Hartree contribution. Here, we eliminate this effect by fixing the fillings of the d and the p orbitals to constant values $\sum_{\si} \langle n^{d}_{\si}\rangle = 1.1$ and $\sum_{\si} \langle n^{p}_{\si}\rangle = 1.9$. 
Without the Hartree terms, even weak intershell interactions significantly change the results, as can be seen in panel (a) of FIG. \ref{fig:MIT} where they correspond to only 10\% of the d-d interaction. Further increasing $U^{dp}$ and $U^{pp}$ to 25\% (panel (b)) and 50\% (panel (c)) of $U^{dd}$ leads to a localization of the conduction electrons, thus inducing a metal-insulator phase transition. \par
Comparing the different approximate schemes allows us to evaluate their performance for different intershell interaction strengths. As it can be anticipated from the perturbative character of the Hartree-Fock approximation, it succeeds rather well in approximating the effect of weak $U^{dp}=U^{pp}$, slightly outperforming the slave rotor method (at least in the low frequency regime -- a closer inspection of the high-frequency tail reveals a better performance of the slave-rotor methods).
The shell folding scheme -- generally presuming a reduction of the effective interaction strength --  fails to predict the direction of the shift even at weak intershell interaction strengths. \par
The strength of our new slave rotor method becomes visible at stronger intershell interactions. In panels (b) and (c) of FIG. \ref{fig:MIT}, the ever larger values of $U^{dp}$ and $U^{pp}$ push the system into the insulating phase, while the shell folding and Hartree-Fock approximations predict increasing spectral weight at the Fermi level. Only the slave rotor method reproduces the metal-insulator transition, giving even better results at stronger interactions. The fact that the slave rotor method is closer to the direct DMFT results at intermediate to strong interactions is reminiscent of analogous observations within the standard Slave Rotor method for the orbitally degenerate Hubbard model\cite{SR_meanfield}. \par
The performance of the slave rotor technique at weak interaction strengths could be optimized by discarding rotor states that are \`a priori unphysical, such as those with angular momentum $Q^{d} + Q^{p} > 2(N^{d} + N^{p})$ ($N^{d}$ and $N^{p}$ being the number of d and p orbitals, respectively; the factor two stems from the spin d.o.f.) and $Q^{d} + Q^{p}<0$. Such an implementation has been suggested in Ref. \onlinecite{Generalized_SP} for the degenerate multi-orbital Hubbard model. In the case of our problem, such an approach would promise slightly better results in the case of commensurate fillings. For incommensurate fillings, however, this technique is problematic since preserving the correct non-interacting limit comes at the cost of violation of the fermionic anti-commutation relations, and vice versa.  
\begin{figure}
\includegraphics[width=0.48\textwidth]{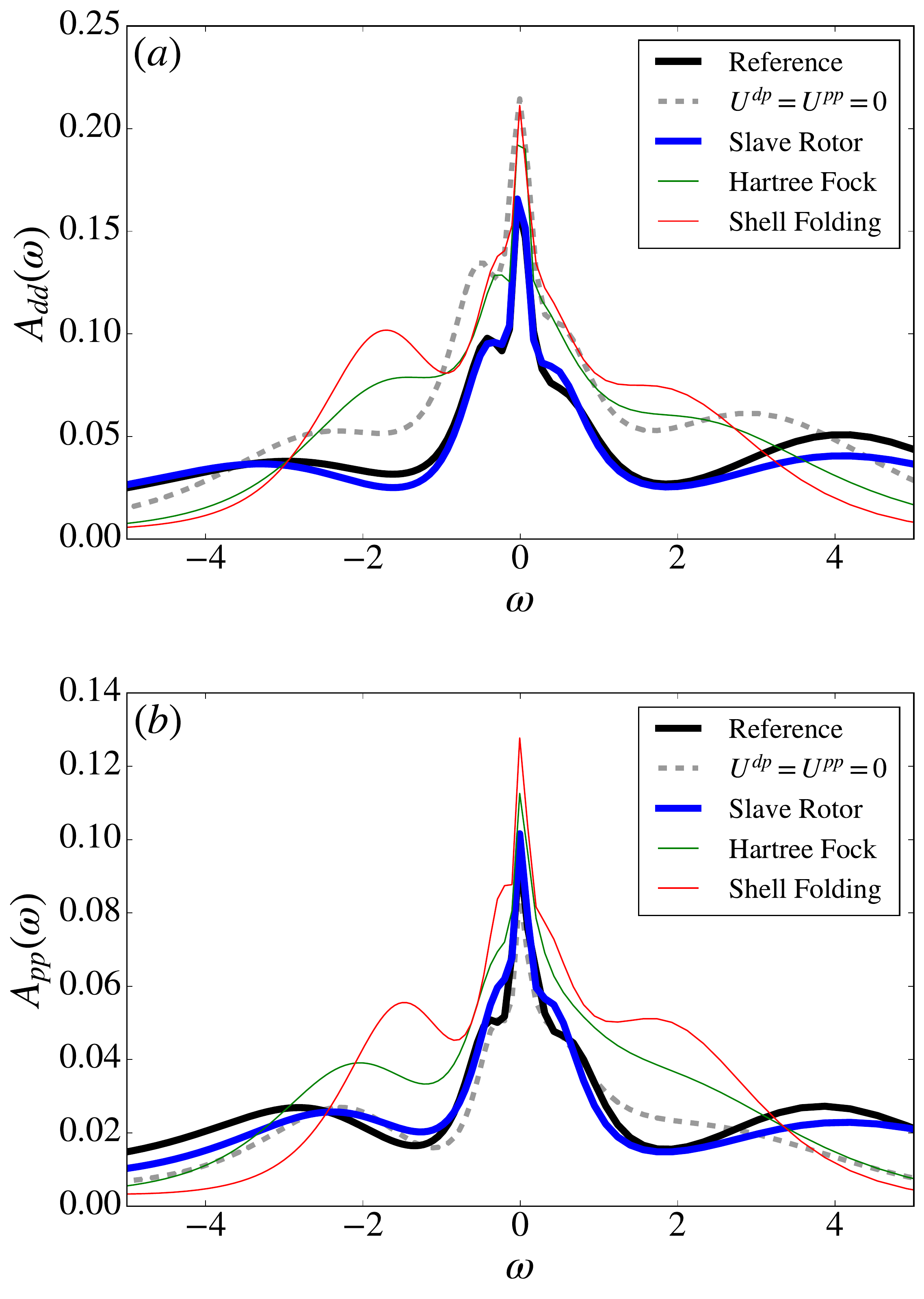}
\caption{Spectral functions of the d orbitals (a) and the p orbitals (b), comparing the QMC results for the full Hamiltonian \eqref{eq:test_model} (reference) to results obtained from different approximate techniques to treat the intershell interactions: 1) Setting $U^{dp}=U^{pp}=0$, 2) our Slave Rotor scheme, 3) Hartree-Fock and 4) Shell Folding. $U^{dd}=10$, $U^{dp}=U^{pp}=0.5U^{dd}$, $\Delta=0$, $V=7$, $\beta=32$. All spectral functions were obtained from analytic continuation of the corresponding Green's functions on the Matsubara axis by using the maximum entropy method\cite{maxent_Gull}.}\label{fig:Spectral1}
\end{figure}
\subsubsection{Spectral properties -- Results}
FIG. \ref{fig:Spectral1} shows the spectral functions of the d and the p orbitals, comparing results from different methods. The parameters $\Delta$, $V$ and $\beta$ are equivalent to those in FIG. \ref{fig:corr_increase_decrease} (b), while $U^{dd}=10$, $U^{dp}=U^{pp}=0.5U^{dd}$ corresponds to a more strongly correlated regime. The results from all different calculations show sharp quasi-particle peaks at the Fermi level, both for the d and the p orbital, confirming the metallic character of the solution. Comparing panels (a) and (b), one remarks a strong similarity between the spectral functions of the two orbitals. This is due to the strong on-site hybridization $V=7$ and the weak charge-transfer energy $\Delta=0$. While this feature is captured by all techniques, only the results from the slave rotor approach remains close the reference curve at all energy scales. Remarkably, the shell folding and Hartree-Fock methods are outperformed by the most trivial technique of simply ignoring the intershell interactions ($U^{dp}=U^{pp}=0$). This indicates, that in the regime under consideration the different effects of intershell interactions (screening from the ligands and increase of the non-local charge-excitation energy) are of a similar magnitude and cancel each other out. Panel (b) of FIG. \ref{fig:Spectral1} demonstrates that the slave rotor technique also provides decent results for the p orbitals. This is remarkable, since the correlations of the p orbitals are solely incorporated via the rotor Green's function. A more extensive study of the spectral properties in different parameter regimes can be found in the Appendix.\par
It shall be noted that care must be taken in the interpretation of the real-frequency results. The spectral functions were obtained from analytic continuation of the Green's functions on the Matsubara axis, which is known to be an ill-defined problem. Within the maximum entropy method\cite{maxent_Gull} which was applied here, the results depend on the specific choice of the default model. In order to allow for an objective comparison of the different methods, we used the same default model for all results corresponding to the same set of parameters. \par
{\color{black} Finally, we shall note that the spectral functions in FIG \ref{fig:Spectral1}, as well as most other spectral functions of metallic character shown in this paper exhibit little humps around the Fermi level. These are reminiscent of the specific form of the DOS of the
2d square lattice, renormalized by the quasi-particle $Z$
factor. Similar structures have been observed in previous works\cite{ayral_spectral}.}

\begin{figure}
\includegraphics[width=0.48\textwidth]{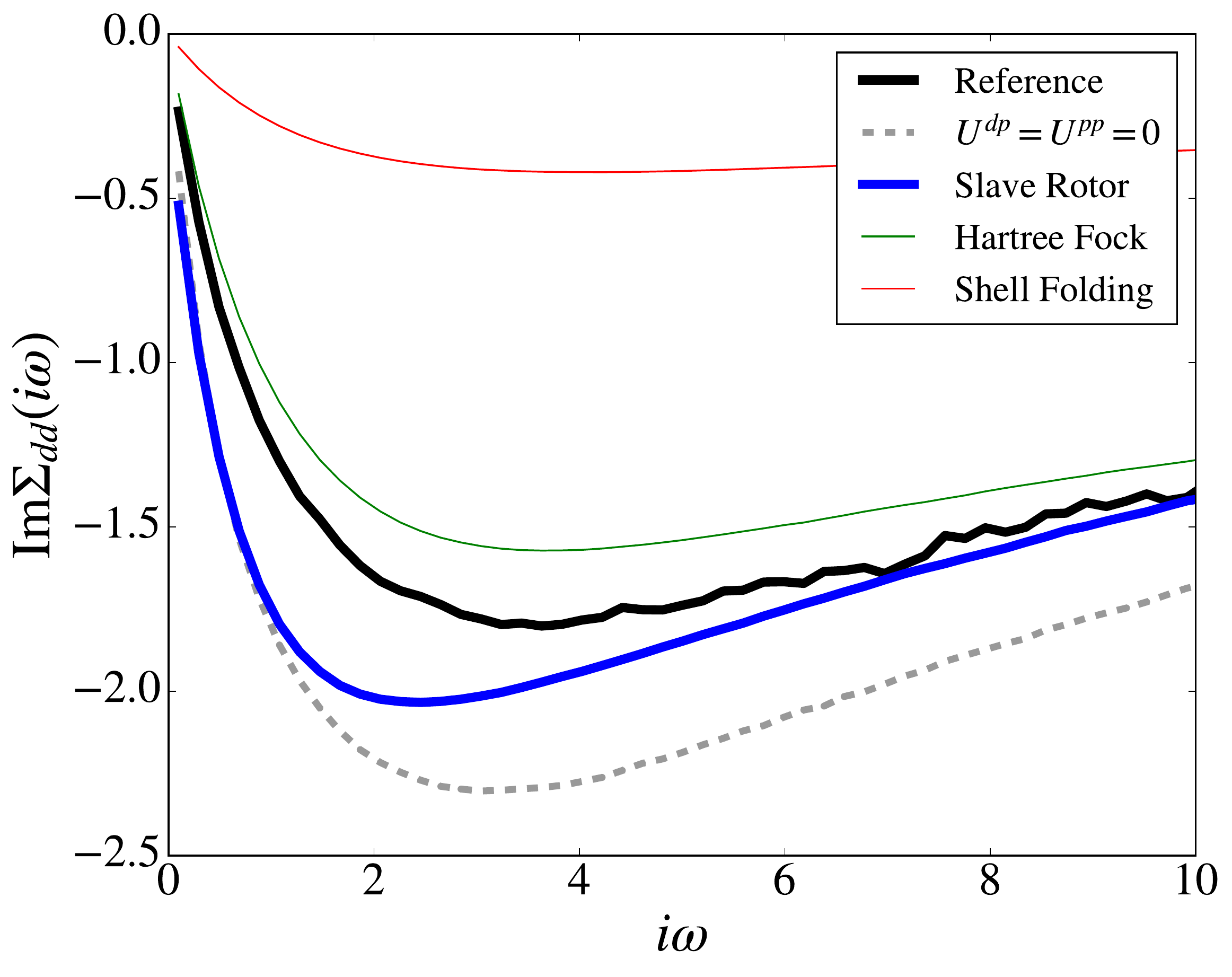}
\caption{Imaginary parts of the d orbital self energy, comparing the QMC results for the full Hamiltonian \eqref{eq:test_model} (reference) corresponding to model Hamiltonian \eqref{eq:test_model} to results obtained from different approximate techniques to treat the intershell interactions: 1) Setting $U^{dp}=U^{pp}=0$, 2) our Slave Rotor scheme, 3) Hartree-Fock and 4) Shell Folding. The total filling is set to $N=2.8$, the model parameters are $U^{dd}=10$, $U^{dp}=U^{pp}=0.5U^{dd}$, $\Delta=0$, $V=3.5$, $\beta=32$.}\label{fig:incommensurate}
\end{figure}
\subsubsection{Incommensurate fillings}
So far, we only considered systems at commensurate fillings. FIG. \ref{fig:incommensurate} shows the imaginary parts of the d orbital self energies for the case of an incommensurate total filling $N=2.8$. \par
Compared to the results corresponding to case of $N=3$, the slave rotor curves show a stronger deviation from the reference curves, which is particularly pronounced at lower frequencies. Earlier works\cite{SR_impurity} already reported limitations of the slave rotor method (in the context of a sigma model approximation to the rotor Hamiltonian) away from half filling.  Here, the increased inaccuracy at low frequencies is of a different origin, and must be attributed to the mean-field decoupling \eqref{eq:SR_meanfield}. At non-integer fillings, it is possible to have low-energy charge excitations; the energy scales thus cease to be well separated. 
On the other hand, the Hartree-Fock method performs remarkably well in the low-frequency regime close to the Fermi level. Only at higher energies, the slave rotor method trumps the Hartree-Fock approximation, providing a good description of the high-frequency tail of the self energies. \\
{\color{black}As a final remark, it shall be noted that we observed the deviations to decrease upon approaching commensurate fillings, making our slave-rotor method well suited for applications in weakly doped regimes.}

\section{Summary and Outlook}\label{sec:discussion}
In summary, we have proposed a novel method that allows for an approximate description of intershell interactions in the context of impurity models with multiple electron shells. Our method is based on the shell folding approximation, which we have extended by taking into account the effect of fluctuations of the total local electron charge. This is achieved using the slave rotor technique, by introducing a rotor variable that represents the fluctuations of the total local charge. \par
Beyond the formal derivation of our technique, we have suggested and tested a practical implementation of our method, which relies on a simple mean-field decoupling of the rotor and fermionic degrees of freedom.
In the case of a commensurate filling $\langle\hat{N}\rangle=3$, we have found that the slave rotor method reproduces the different effects which intershell interactions can have on the electronic correlations, clearly outperforming the shell folding and the Hartree-Fock approximations. We further show that our method is capable of faithfully reproducing the scenario of an intershell interaction-driven metal-insulator transition.
The quality of the slave rotor method is reduced in the case of incommensurate fillings. In this case, deviations from the reference results are found in the low frequency domain. Remarkably, however, the high frequency features are still reproduced accurately. \par
In this paper, we applied our method to a minimal model system. Its main merit, however, lies in its potential to allow calculations of more complex, realistic systems:\par
Potential applications of our method include all late transition metal oxides, where there is significant metal-to-ligand charge transfer. Even though intershell interactions in such compounds can be significant\cite{shell_folding,Hansmann_2014}, they are usually ignored, or only treated on the Hartree-level. Here, our slave rotor method opens new routes for more realistic calculations.\par
One application -- and natural extension to the model Hamiltonian considered here -- might be the three-band Emery model\cite{Emery_model,Andersen_Emery} for cuprate systems. Theoretical studies on this model emphasized the importance of $U^{dp}$ in stabilizing the charge-transfer insulating state\cite{Hansmann_2014} (in that work, the intershell interaction was, however, only treated on a Hartree level).  
More generally, our approach bears promise for treating late transition metal oxides, where d-p fluctuations are expected to play a role. An example would be NiO where the intershell interaction was calculated to be $U^{dp}=2.2$ eV, which exceeds 25\% of the d-d interaction\cite{shell_folding}. Including the corresponding terms might prove important to improve upon the -- hitherto rather unsatisfactory\cite{NiO_dc} -- agreement between theoretical and experimental spectral properties.
Similarly, our slave rotor method might be of use to further investigate the physics of other nickelate systems. 
Previous model studies were restricted to treating the ligand states as uncorrelated\cite{dp_millis,dp_millis_medici} -- or did not include them at all\cite{d_hansmann}. \par
In the context of realisitic materials calculations, it is of paramount interest to access real-frequency properties of the quantum systems such as the electronic spectral function. Here, the pathological problem of the analytic continuation could be circumvented by using methods like exact diagonalization or the numerical renormalization group\cite{Wilson_NRG}, which would allow direct access to 
 Green's function. Since the rotor Hamiltonian \eqref{eq:H_theta} can be easily diagonalized, the ``composite'', physical spectrum could then be obtained directly.\par
The accuracy of the slave rotor method might be further improved by considering corrections beyond the mean-field approximation, which was applied to decouple the rotor and pseudo fermionic operators. This should also lead to an improved description in the case of incommensurate fillings. Another route might be to modify the rotor Green's function to incorporate information about the anticipated low-energy behavior from analytic considerations. A similar approach has been successfully applied in the description of satellite features due to dynamic interactions\cite{DALA_Casula}.

\section{Acknowledgements}
We thank Serge Florens and Mark van Schilfgaarde, as well as Anna Galler, Sumanta Bhandary and Christoph Glatz for interesting and stimulating discussions. This work was supported by the European Research Council (Consolidator Grant No. 617196 CORRELMAT) and supercomputing time 
at IDRIS/GENCI Orsay (Project No. t2020091393).
We thank the computer team at CPHT for support.

\appendix

\section{Non-local charge excitation in the atomic limit}\label{sec:atomic_limit_energy}
The results of the numeric simulations for the d-p model showed, that intershell interactions can have varying effects on the electronic correlations, depending on the parameter regimes under consideration. While screening from the ligand orbitals causes a decrease in correlation, there is an increased energy cost associated with non-local charge excitations, which can lead to an increase of correlations. \par
\begin{figure}
\includegraphics[width=0.48\textwidth]{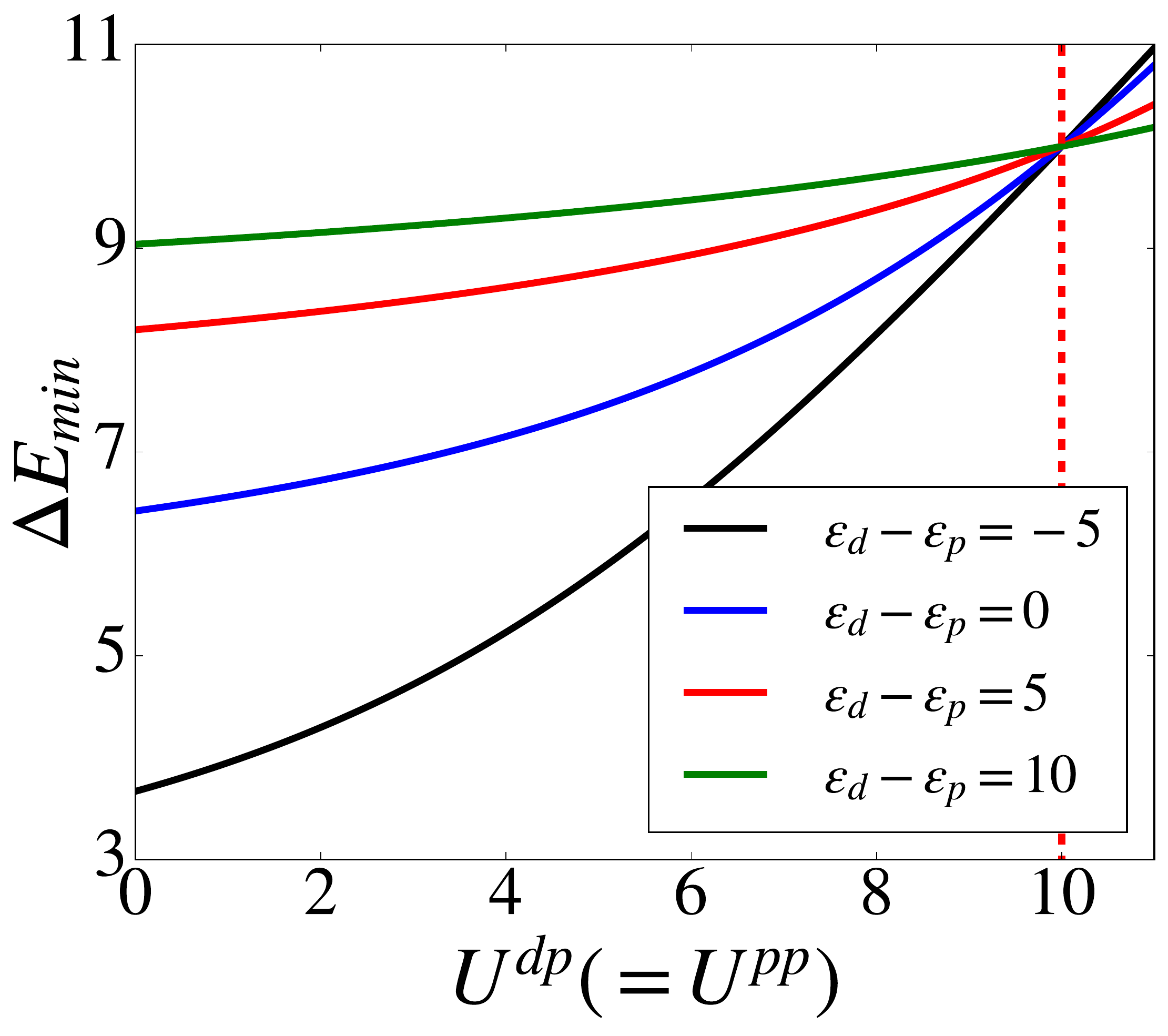}
\caption{Energy cost for a non-local charge transfer in the atomic limit (the lattice model corresponds to Hamiltonian \eqref{eq:test_model} -- see Fig. \ref{fig:Udp_model} -- with $t^{dp}=0$), with N=3 electrons per site. Various lines show the results for different charge-transfer enegies; the d orbital interaction is fixed to $U^{dd}=10$ and the on-site hybridization is $V=3.5$. }\label{fig:atomic_limit_energy}
\end{figure}

The latter effect is easily demonstrated in the atomic limit: Fig. \ref{fig:atomic_limit_energy} shows the energy cost of moving an electron from one lattice site to another (the corresponding system is illustrated in Fig. \ref{fig:Udp_model}), when the initial filling per site is $N=3$ and the sites are isolated from each other $t^{dp}=0$. The figure clearly shows that, independently of the charge-transfer energy, intershell iteractions increase the energy cost for displacing an electron. It must be noted that this is only true if the local hybridization $V\neq 0$. Otherwise, the result is independent of $U^{dp}$ and $U^{pp}$.

\section{Benchmark test of the slave rotor technique}\label{sec:A1}
In this section we present results that allow for comparison of the slave rotor, shell folding and Hartree-Fock approximation in different parameter regimes. As before, we consider the two orbital model which we introduced in Section \ref{sec:the_test_model}; the curves from the approximate methods are again benchmarked against those from the exact quantum Monte Carlo calculation (reference) and the results without $U^{dp}$ and $U^{pp}$. All figures in this section share the same structure: Panels (a) and (b) show the d- and p orbital Green's functions on the imaginary time axis; the insets show the logarithm of the (absolute value of the) same quantities, in order to improve the distinction of the different curves. Panels (c) and (b) show the spectral functions of the d- and p orbitals, obtained from analytic continuation using the maximum entropy method\cite{maxent_Gull}. Care must be taking when drawing conclusions from the comparison of these spectral functions, since the results of the analytic continuation algorithm is sensitive to the input parameters of the routine. Finally, panels (e) and (f) focus on the imaginary parts of the Green's function and self-energy of the d orbital, which providing insights into the physics of the strongly correlated subspace. All of the following results were obtained at inverse temperature $\beta=32$; as before, all parameters are given in units of the off-diagonal inter-site hopping $t^{dp}$. \par
\begin{figure*}
\large{\textbf{Parameters: $N=3$, $U^{dd}=8.75$, $U^{dp}=U^{pp}=0.5U^{dd}$, $V=3.5$ , $\Delta=0$.}}\par\medskip
\includegraphics[width=0.9\textwidth]{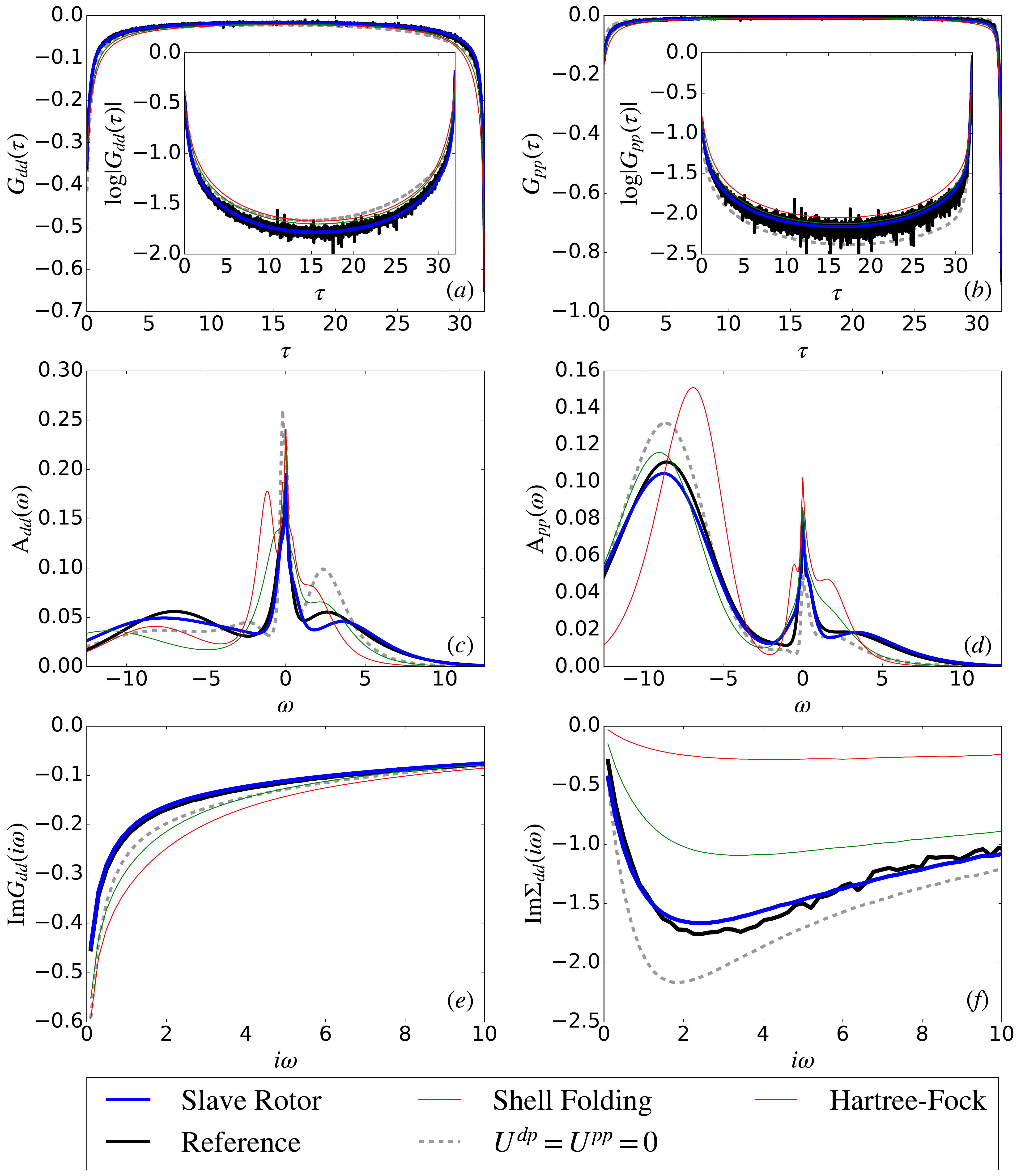}
\caption{Benchmarking different approximative methods to treat the intershell interaction in comparison to the Monte Carlo results for the full Hamiltonian Eq. \eqref{eq:test_model} 
(black curve; ``Reference''). 
Panels (a) and (b) show the imaginary time Green's functions for the d and p orbital, respectively; panels (c) and (d) show the spectral functions for the d and p orbitals (respectively), obtained from the maximum entropy analytic continuation; panels (e) and (f) show the imaginary part of the Green's function and self-energy (respectively) for the strongly correlated d orbital. $U^{dd}=8.75$, $U^{dp}=U^{pp}=0.5U^{dd}$, $\epsilon^{d}=0$, $\epsilon^{p}=0$, $V=3.5$, with total filling $N=3$ and $\beta=32$. }\label{fig:N3V35D0_1}
\end{figure*}
\begin{figure*}
\large{\textbf{Parameters: $N=3$, $U^{dd}=10$, $U^{dp}=U^{pp}=0.5U^{dd}$, $V=3.5$ , $\Delta=0$.}}\par\medskip
\includegraphics[width=0.9\textwidth]{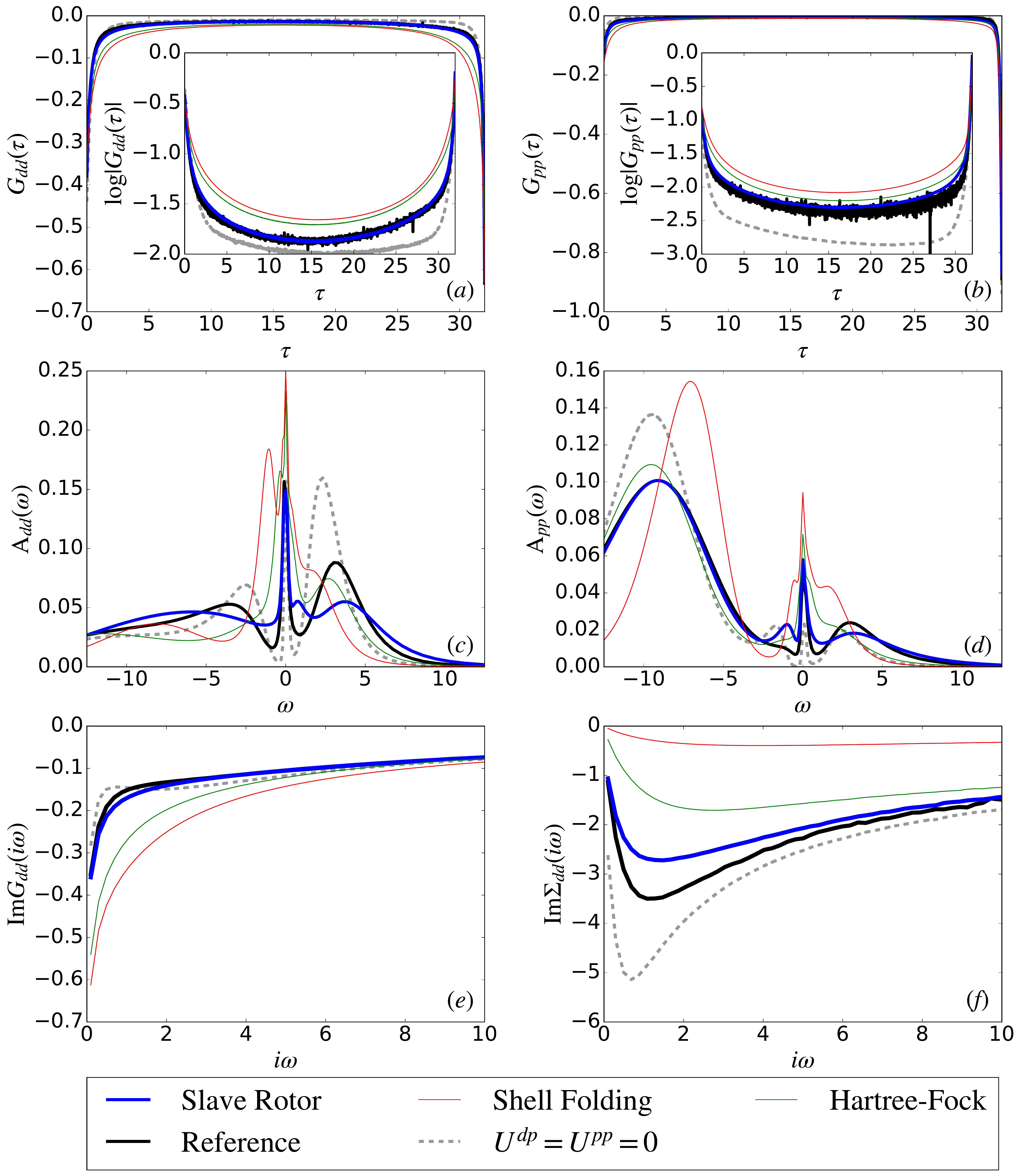}
\caption{Benchmarking different approximative methods to treat the intershell interaction in comparison to the Monte Carlo results for the full Hamiltonian Eq. \eqref{eq:test_model} 
(black curve; ``Reference''). 
Panels (a) and (b) show the imaginary time Green's functions for the d and p orbital, respectively; panels (c) and (d) show the spectral functions for the d and p orbitals (respectively), obtained from the maximum entropy analytic continuation; panels (e) and (f) show the imaginary part of the Green's function and self-energy (respectively) for the strongly correlated d orbital. $U^{dd}=10$, $U^{dp}=U^{pp}=0.5U^{dd}$, $\epsilon^{d}=0$, $\epsilon^{p}=0$, $V=3.5$, with total filling $N=3$ and $\beta=32$. }\label{fig:N3V35D0_2}
\end{figure*}

Fig. \ref{fig:N3V35D0_1} and \ref{fig:N3V35D0_2} were obtained at commensurate filling $\langle N\rangle =3$, on-site hybridization $V=3.5$ and bare charge-transfer energy $\Delta=0$, for $U^{dd}=8.75$ and $U^{dd}=10$, respectively. In both cases, the additional interactions $U^{dp}=U^{pp}=0.5U^{dd}$ decrease the correlations. In Fig. \ref{fig:N3V35D0_2}, the system is in the vicinity of the metal-insulator transition, and deviations from the reference solution are stronger.  \par
\begin{figure*}
\large{\textbf{Parameters: $N=3$, $U^{dd}=12.5$, $U^{dp}=U^{pp}=0.5U^{dd}$, $V=3.5$ , $\Delta=-5$.}}\par\medskip
\includegraphics[width=0.9\textwidth]{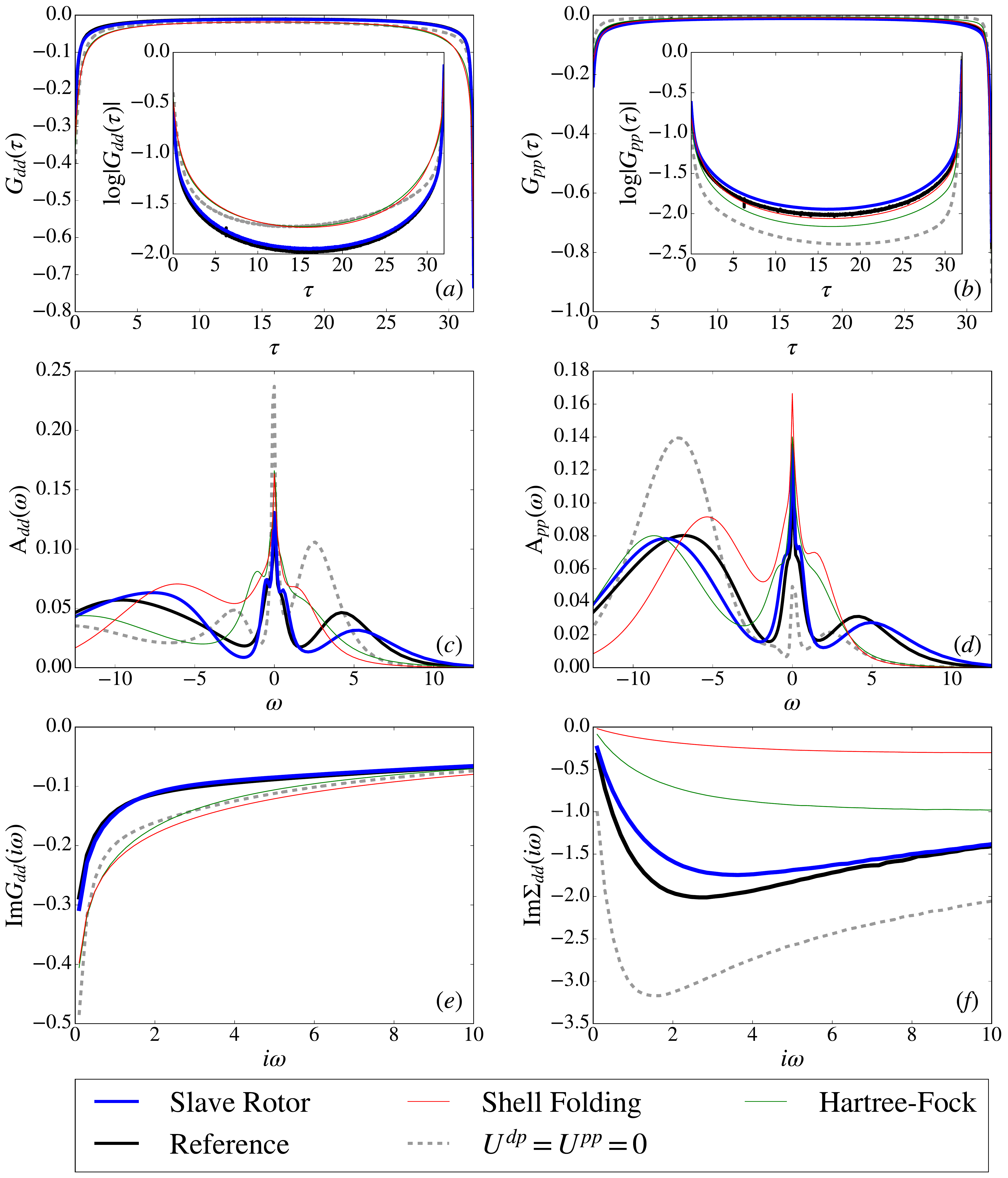}
\caption{Benchmarking different approximative methods to treat the intershell interaction in comparison to the Monte Carlo results for the full Hamiltonian Eq. \eqref{eq:test_model} 
(black curve; ``Reference''). 
Panels (a) and (b) show the imaginary time Green's functions for the d and p orbital, respectively; panels (c) and (d) show the spectral functions for the d and p orbitals (respectively), obtained from the maximum entropy analytic continuation; panels (e) and (f) show the imaginary part of the Green's function and self-energy (respectively) for the strongly correlated d orbital. $U^{dd}=12.5$, $U^{dp}=U^{pp}=0.5U^{dd}$, $\epsilon^{d}=0$, $\epsilon^{p}=5$, $V=3.5$, with total filling $N=3$ and $\beta=32$. }\label{fig:N3V35D2_1}
\end{figure*}

Fig. \ref{fig:N3V35D2_1} was obtained at commensurate filling $\langle N\rangle =3$ and on-site hybridization $V=3.5$ however the bare charge-transfer energy and the d-d interaction are changed to $\Delta=-5$ and $U^{dd}=12.5$, respectively. The increase of the bare p orbital energy leads to an enhanced p-d charge transfer, which amplifies the effect of the intershell interactions. \par
\begin{figure*}
\large{\textbf{Parameters: $N=3$, $U^{dd}=10$, $U^{dp}=U^{pp}=0.5U^{dd}$, $V=7$ , $\Delta=0$.}}\par\medskip
\includegraphics[width=0.9\textwidth]{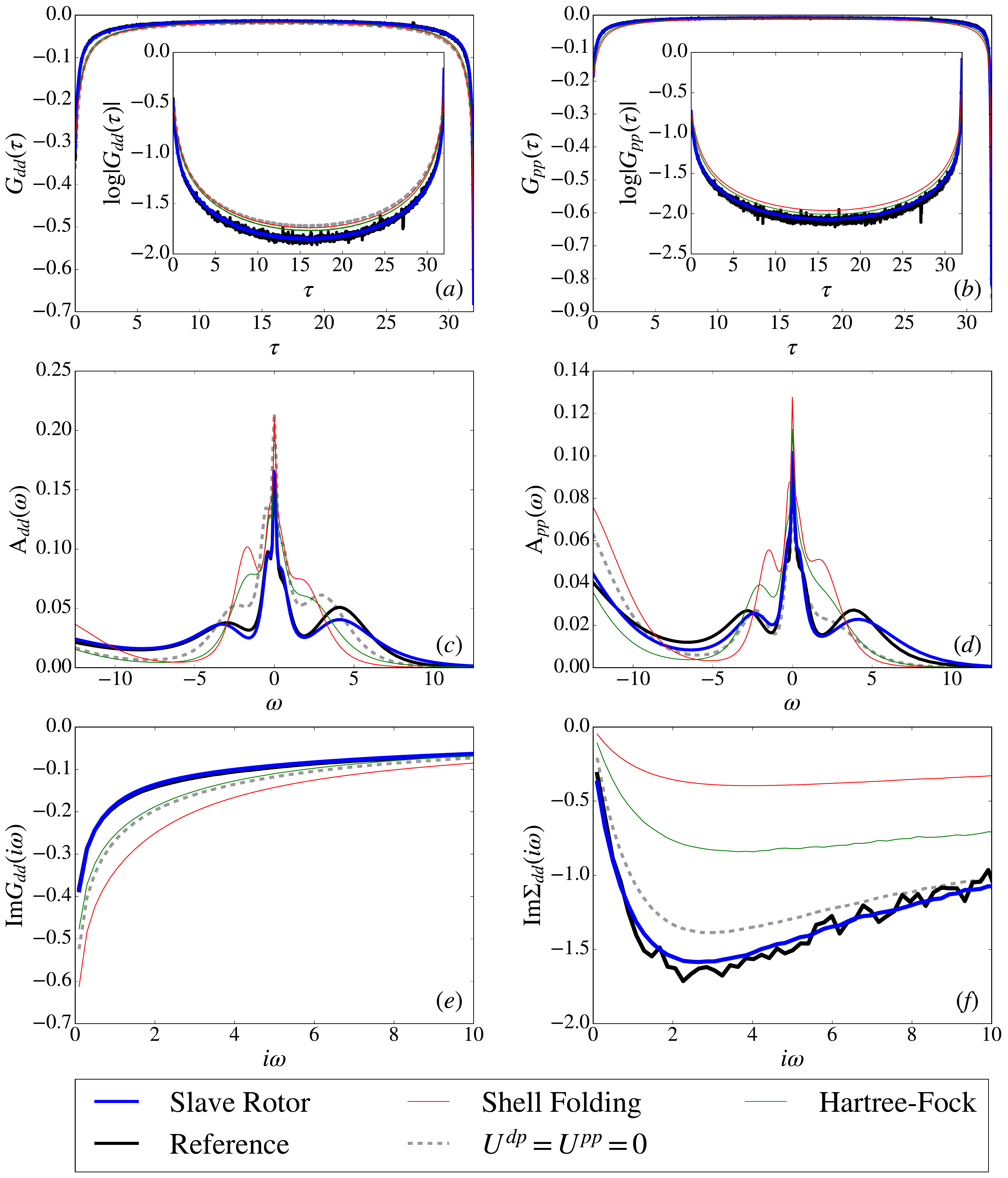}
\caption{Benchmarking different approximative methods to treat the intershell interaction in comparison to the Monte Carlo results for the full Hamiltonian Eq. \eqref{eq:test_model} 
(black curve; ``Reference''). 
Panels (a) and (b) show the imaginary time Green's functions for the d and p orbital, respectively; panels (c) and (d) show the spectral functions for the d and p orbitals (respectively), obtained from the maximum entropy analytic continuation; panels (e) and (f) show the imaginary part of the Green's function and self-energy (respectively) for the strongly correlated d orbital. $U^{dd}=10$, $U^{dp}=U^{pp}=0.5U^{dd}$, $\epsilon^{d}=0$, $\epsilon^{p}=0$, $V=7$, with total filling $N=3$ and $\beta=32$. }\label{fig:N3V7D0_1}
\end{figure*}
\begin{figure*}
\large{\textbf{Parameters: $N=3$, $U^{dd}=15$, $U^{dp}=U^{pp}=0.5U^{dd}$, $V=7$ , $\Delta=0$.}}\par\medskip
\includegraphics[width=0.9\textwidth]{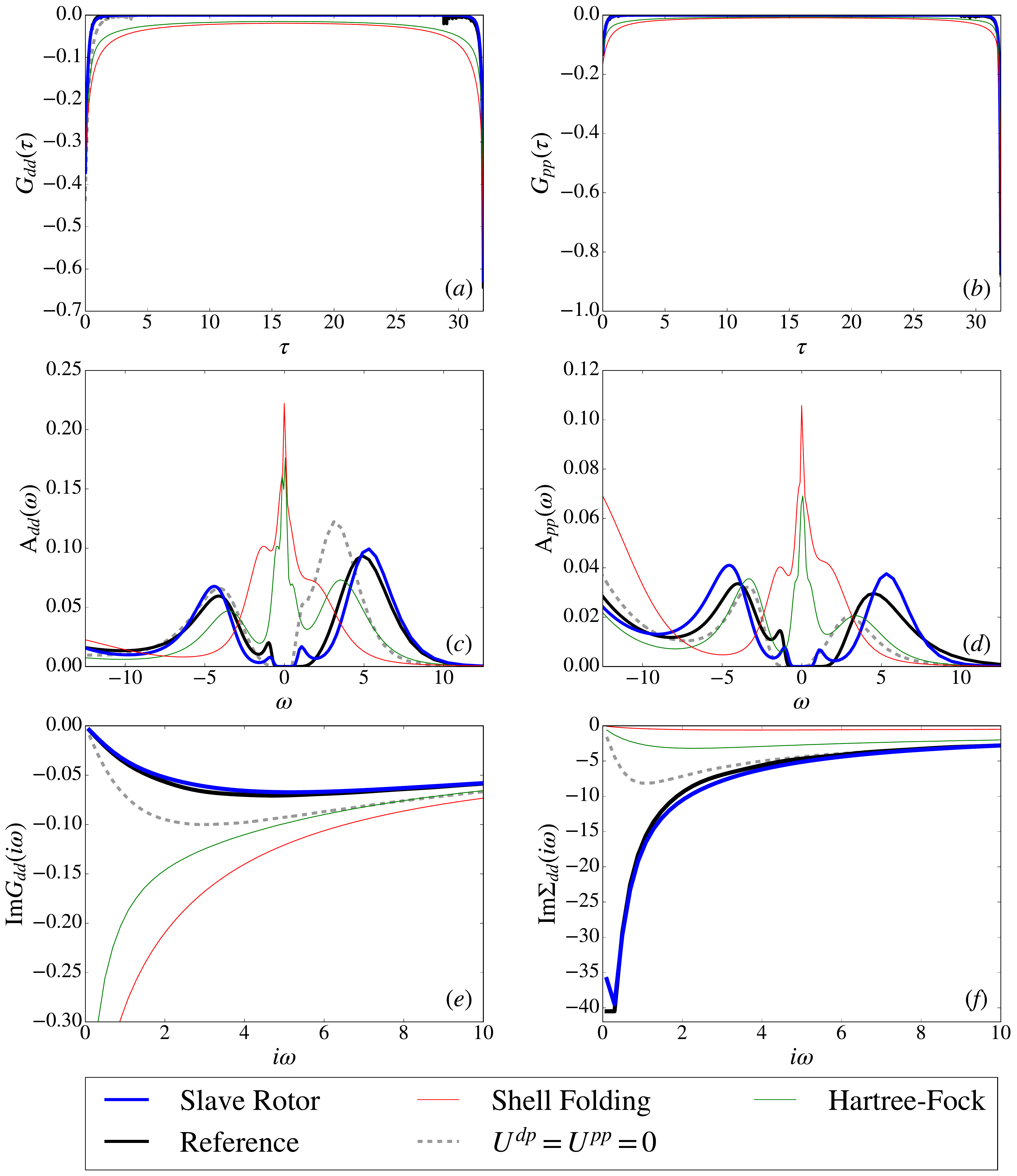}
\caption{Benchmarking different approximative methods to treat the intershell interaction in comparison to the Monte Carlo results for the full Hamiltonian Eq. \eqref{eq:test_model} 
(black curve; ``Reference''). 
Panels (a) and (b) show the imaginary time Green's functions for the d and p orbital, respectively; panels (c) and (d) show the spectral functions for the d and p orbitals (respectively), obtained from the maximum entropy analytic continuation; panels (e) and (f) show the imaginary part of the Green's function and self-energy (respectively) for the strongly correlated d orbital. $U^{dd}=15$, $U^{dp}=U^{pp}=0.5U^{dd}$, $\epsilon^{d}=0$, $\epsilon^{p}=0$, $V=7$, with total filling $N=3$ and $\beta=32$. }\label{fig:N3V7D0_2}
\end{figure*}

Fig. \ref{fig:N3V7D0_1} and \ref{fig:N3V7D0_2} show the results for commensurate filling $\langle N\rangle =3$, on-site hybridization $V=7$ and bare charge-transfer energy $\Delta=0$, for $U^{dd}=10$ and $U^{dd}=15$, respectively. In the first case, the system is in the metallic regime and turning on $U^{dp}=U^{pp}=0.5U^{dd}$ leads to an increase in correlations. In Fig. \ref{fig:N3V7D0_2}, the system is a charge-transfer insulator. Here, the additional interactions result in a broadening of the spectral gap. The results from the shell folding and the Hartree-Fock approximations are completely off the reference results, even predicting metallic solutions. The result from the slave rotor method, on the other hand, is in good agreement with the reference.\par
\begin{figure*}
\large{\textbf{Parameters: $N=2.8$, $U^{dd}=10$, $U^{dp}=U^{pp}=0.5U^{dd}$, $V=3.5$ , $\Delta=0$.}}\par\medskip
\includegraphics[width=0.9\textwidth]{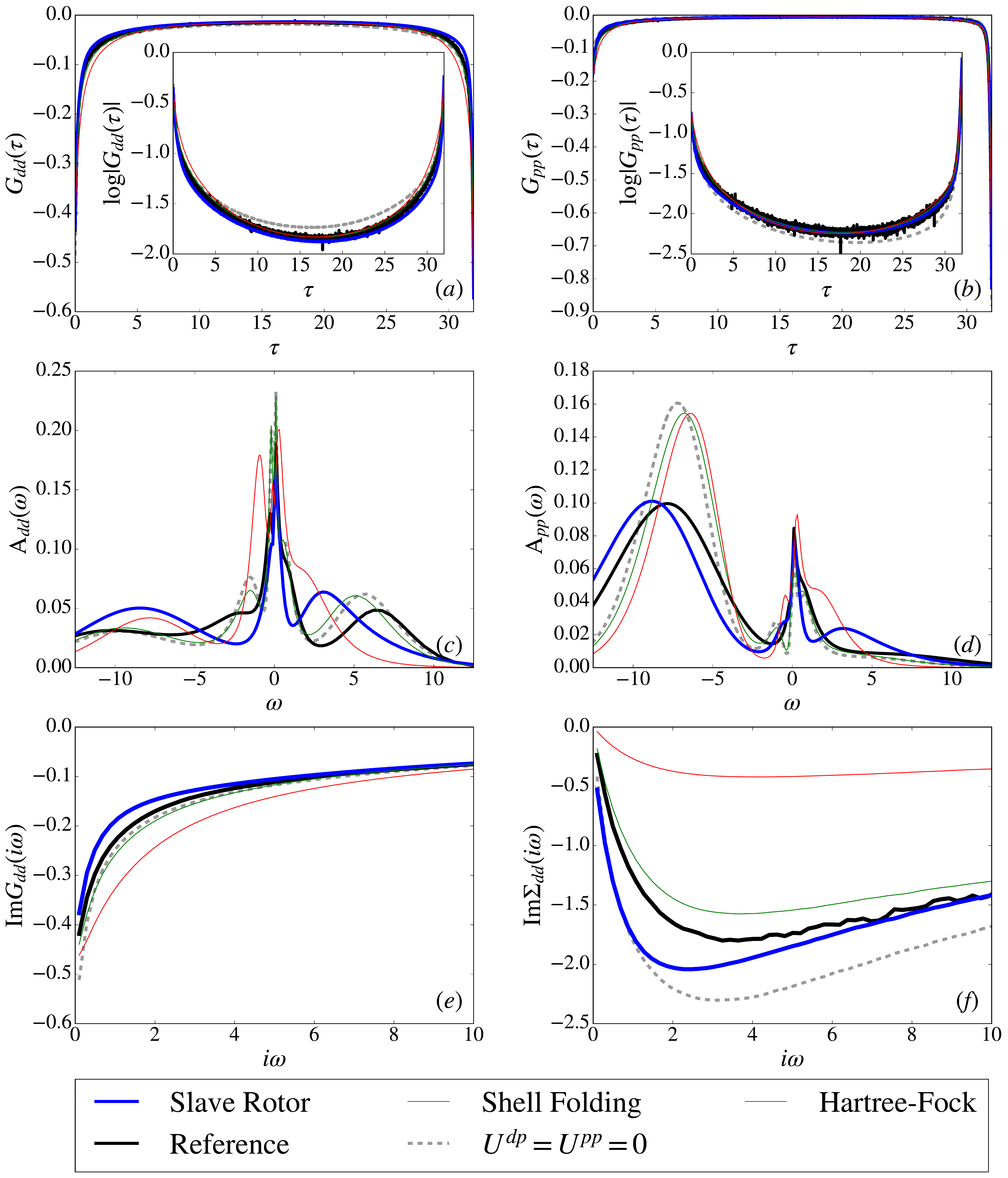}
\caption{Benchmarking different approximative methods to treat the intershell interaction in comparison to the Monte Carlo results for the full Hamiltonian Eq. \eqref{eq:test_model} 
(black curve; ``Reference''). 
Panels (a) and (b) show the imaginary time Green's functions for the d and p orbital, respectively; panels (c) and (d) show the spectral functions for the d and p orbitals (respectively), obtained from the maximum entropy analytic continuation; panels (e) and (f) show the imaginary part of the Green's function and self-energy (respectively) for the strongly correlated d orbital. $U^{dd}=10$, $U^{dp}=U^{pp}=0.5U^{dd}$, $\epsilon^{d}=0$, $\epsilon^{p}=0$, $V=3.5$, with total filling $N=2.8$ and $\beta=32$. }\label{fig:N28V35D0_1}
\end{figure*}

Fig. \ref{fig:N28V35D0_1} finally shows the results at an incommensurate filling $\langle N\rangle =2.8$ with an on-site hybridization $V=3.5$, bare charge-transfer energy $\Delta=0$ and  $U^{dd}=10$. As explained above, the slave rotor solution shows a stronger deviation at low frequencies; the Hartree-Fock approximation, on the other hand is rather good in this regime. Only at higher frequencies, the slave rotor technique trumps the other methods.

\nocite{*}

\begin{thebibliography}{64}%
\makeatletter
\providecommand \@ifxundefined [1]{%
 \@ifx{#1\undefined}
}%
\providecommand \@ifnum [1]{%
 \ifnum #1\expandafter \@firstoftwo
 \else \expandafter \@secondoftwo
 \fi
}%
\providecommand \@ifx [1]{%
 \ifx #1\expandafter \@firstoftwo
 \else \expandafter \@secondoftwo
 \fi
}%
\providecommand \natexlab [1]{#1}%
\providecommand \enquote  [1]{``#1''}%
\providecommand \bibnamefont  [1]{#1}%
\providecommand \bibfnamefont [1]{#1}%
\providecommand \citenamefont [1]{#1}%
\providecommand \href@noop [0]{\@secondoftwo}%
\providecommand \href [0]{\begingroup \@sanitize@url \@href}%
\providecommand \@href[1]{\@@startlink{#1}\@@href}%
\providecommand \@@href[1]{\endgroup#1\@@endlink}%
\providecommand \@sanitize@url [0]{\catcode `\\12\catcode `\$12\catcode
  `\&12\catcode `\#12\catcode `\^12\catcode `\_12\catcode `\%12\relax}%
\providecommand \@@startlink[1]{}%
\providecommand \@@endlink[0]{}%
\providecommand \url  [0]{\begingroup\@sanitize@url \@url }%
\providecommand \@url [1]{\endgroup\@href {#1}{\urlprefix }}%
\providecommand \urlprefix  [0]{URL }%
\providecommand \Eprint [0]{\href }%
\providecommand \doibase [0]{http://dx.doi.org/}%
\providecommand \selectlanguage [0]{\@gobble}%
\providecommand \bibinfo  [0]{\@secondoftwo}%
\providecommand \bibfield  [0]{\@secondoftwo}%
\providecommand \translation [1]{[#1]}%
\providecommand \BibitemOpen [0]{}%
\providecommand \bibitemStop [0]{}%
\providecommand \bibitemNoStop [0]{.\EOS\space}%
\providecommand \EOS [0]{\spacefactor3000\relax}%
\providecommand \BibitemShut  [1]{\csname bibitem#1\endcsname}%
\let\auto@bib@innerbib\@empty
\bibitem [{\citenamefont {de~Boer}\ and\ \citenamefont
  {Verwey}(1937)}]{Boer_1937}%
  \BibitemOpen
  \bibfield  {author} {\bibinfo {author} {\bibfnamefont {J.~H.}\ \bibnamefont
  {de~Boer}}\ and\ \bibinfo {author} {\bibfnamefont {E.~J.~W.}\ \bibnamefont
  {Verwey}},\ }\href {\doibase 10.1088/0959-5309/49/4s/307} {\bibfield
  {journal} {\bibinfo  {journal} {Proceedings of the Physical Society}\
  }\textbf {\bibinfo {volume} {49}},\ \bibinfo {pages} {59} (\bibinfo {year}
  {1937})}\BibitemShut {NoStop}%
\bibitem [{\citenamefont {Mott}\ and\ \citenamefont
  {Peierls}(1937)}]{Mott_1937}%
  \BibitemOpen
  \bibfield  {author} {\bibinfo {author} {\bibfnamefont {N.~F.}\ \bibnamefont
  {Mott}}\ and\ \bibinfo {author} {\bibfnamefont {R.}~\bibnamefont {Peierls}},\
  }\href {\doibase 10.1088/0959-5309/49/4s/308} {\bibfield  {journal} {\bibinfo
   {journal} {Proceedings of the Physical Society}\ }\textbf {\bibinfo {volume}
  {49}},\ \bibinfo {pages} {72} (\bibinfo {year} {1937})}\BibitemShut {NoStop}%
\bibitem [{\citenamefont {Brinkman}\ and\ \citenamefont
  {Rice}(1970)}]{BrinkmanRice}%
  \BibitemOpen
  \bibfield  {author} {\bibinfo {author} {\bibfnamefont {W.~F.}\ \bibnamefont
  {Brinkman}}\ and\ \bibinfo {author} {\bibfnamefont {T.~M.}\ \bibnamefont
  {Rice}},\ }\href {\doibase 10.1103/PhysRevB.2.4302} {\bibfield  {journal}
  {\bibinfo  {journal} {Phys. Rev. B}\ }\textbf {\bibinfo {volume} {2}},\
  \bibinfo {pages} {4302} (\bibinfo {year} {1970})}\BibitemShut {NoStop}%
\bibitem [{\citenamefont {Rozenberg}\ \emph {et~al.}(1999)\citenamefont
  {Rozenberg}, \citenamefont {Chitra},\ and\ \citenamefont
  {Kotliar}}]{MIT_DMFT_RozenbergKotliar}%
  \BibitemOpen
  \bibfield  {author} {\bibinfo {author} {\bibfnamefont {M.~J.}\ \bibnamefont
  {Rozenberg}}, \bibinfo {author} {\bibfnamefont {R.}~\bibnamefont {Chitra}}, \
  and\ \bibinfo {author} {\bibfnamefont {G.}~\bibnamefont {Kotliar}},\ }\href
  {\doibase 10.1103/PhysRevLett.83.3498} {\bibfield  {journal} {\bibinfo
  {journal} {Phys. Rev. Lett.}\ }\textbf {\bibinfo {volume} {83}},\ \bibinfo
  {pages} {3498} (\bibinfo {year} {1999})}\BibitemShut {NoStop}%
\bibitem [{\citenamefont {Imada}\ \emph {et~al.}(1998)\citenamefont {Imada},
  \citenamefont {Fujimori},\ and\ \citenamefont {Tokura}}]{ImadaReview}%
  \BibitemOpen
  \bibfield  {author} {\bibinfo {author} {\bibfnamefont {M.}~\bibnamefont
  {Imada}}, \bibinfo {author} {\bibfnamefont {A.}~\bibnamefont {Fujimori}}, \
  and\ \bibinfo {author} {\bibfnamefont {Y.}~\bibnamefont {Tokura}},\ }\href
  {\doibase 10.1103/RevModPhys.70.1039} {\bibfield  {journal} {\bibinfo
  {journal} {Rev. Mod. Phys.}\ }\textbf {\bibinfo {volume} {70}},\ \bibinfo
  {pages} {1039} (\bibinfo {year} {1998})}\BibitemShut {NoStop}%
\bibitem [{\citenamefont {Hubbard}\ and\ \citenamefont
  {Flowers}(1963)}]{Hubbard_model}%
  \BibitemOpen
  \bibfield  {author} {\bibinfo {author} {\bibfnamefont {J.}~\bibnamefont
  {Hubbard}}\ and\ \bibinfo {author} {\bibfnamefont {B.~H.}\ \bibnamefont
  {Flowers}},\ }\href {\doibase 10.1098/rspa.1963.0204} {\bibfield  {journal}
  {\bibinfo  {journal} {Proceedings of the Royal Society of London. Series A.
  Mathematical and Physical Sciences}\ }\textbf {\bibinfo {volume} {276}},\
  \bibinfo {pages} {238} (\bibinfo {year} {1963})}\BibitemShut {NoStop}%
\bibitem [{\citenamefont {Gunnarsson}\ \emph {et~al.}(1989)\citenamefont
  {Gunnarsson}, \citenamefont {Andersen}, \citenamefont {Jepsen},\ and\
  \citenamefont {Zaanen}}]{Gunnarsson1}%
  \BibitemOpen
  \bibfield  {author} {\bibinfo {author} {\bibfnamefont {O.}~\bibnamefont
  {Gunnarsson}}, \bibinfo {author} {\bibfnamefont {O.~K.}\ \bibnamefont
  {Andersen}}, \bibinfo {author} {\bibfnamefont {O.}~\bibnamefont {Jepsen}}, \
  and\ \bibinfo {author} {\bibfnamefont {J.}~\bibnamefont {Zaanen}},\ }\href
  {\doibase 10.1103/PhysRevB.39.1708} {\bibfield  {journal} {\bibinfo
  {journal} {Phys. Rev. B}\ }\textbf {\bibinfo {volume} {39}},\ \bibinfo
  {pages} {1708} (\bibinfo {year} {1989})}\BibitemShut {NoStop}%
\bibitem [{\citenamefont {Anisimov}\ and\ \citenamefont
  {Gunnarsson}(1991)}]{Gunnarsson_Anisimov}%
  \BibitemOpen
  \bibfield  {author} {\bibinfo {author} {\bibfnamefont {V.~I.}\ \bibnamefont
  {Anisimov}}\ and\ \bibinfo {author} {\bibfnamefont {O.}~\bibnamefont
  {Gunnarsson}},\ }\href {\doibase 10.1103/PhysRevB.43.7570} {\bibfield
  {journal} {\bibinfo  {journal} {Phys. Rev. B}\ }\textbf {\bibinfo {volume}
  {43}},\ \bibinfo {pages} {7570} (\bibinfo {year} {1991})}\BibitemShut
  {NoStop}%
\bibitem [{\citenamefont {Anisimov}\ \emph {et~al.}(1991)\citenamefont
  {Anisimov}, \citenamefont {Zaanen},\ and\ \citenamefont
  {Andersen}}]{Anisimov_cLDA}%
  \BibitemOpen
  \bibfield  {author} {\bibinfo {author} {\bibfnamefont {V.~I.}\ \bibnamefont
  {Anisimov}}, \bibinfo {author} {\bibfnamefont {J.}~\bibnamefont {Zaanen}}, \
  and\ \bibinfo {author} {\bibfnamefont {O.~K.}\ \bibnamefont {Andersen}},\
  }\href {\doibase 10.1103/PhysRevB.44.943} {\bibfield  {journal} {\bibinfo
  {journal} {Phys. Rev. B}\ }\textbf {\bibinfo {volume} {44}},\ \bibinfo
  {pages} {943} (\bibinfo {year} {1991})}\BibitemShut {NoStop}%
\bibitem [{\citenamefont {Aryasetiawan}\ \emph {et~al.}(2004)\citenamefont
  {Aryasetiawan}, \citenamefont {Imada}, \citenamefont {Georges}, \citenamefont
  {Kotliar}, \citenamefont {Biermann},\ and\ \citenamefont
  {Lichtenstein}}]{cRPA_Aryasetiawan1}%
  \BibitemOpen
  \bibfield  {author} {\bibinfo {author} {\bibfnamefont {F.}~\bibnamefont
  {Aryasetiawan}}, \bibinfo {author} {\bibfnamefont {M.}~\bibnamefont {Imada}},
  \bibinfo {author} {\bibfnamefont {A.}~\bibnamefont {Georges}}, \bibinfo
  {author} {\bibfnamefont {G.}~\bibnamefont {Kotliar}}, \bibinfo {author}
  {\bibfnamefont {S.}~\bibnamefont {Biermann}}, \ and\ \bibinfo {author}
  {\bibfnamefont {A.~I.}\ \bibnamefont {Lichtenstein}},\ }\href {\doibase
  10.1103/PhysRevB.70.195104} {\bibfield  {journal} {\bibinfo  {journal} {Phys.
  Rev. B}\ }\textbf {\bibinfo {volume} {70}},\ \bibinfo {pages} {195104}
  (\bibinfo {year} {2004})}\BibitemShut {NoStop}%
\bibitem [{\citenamefont {Aryasetiawan}\ \emph {et~al.}(2006)\citenamefont
  {Aryasetiawan}, \citenamefont {Karlsson}, \citenamefont {Jepsen},\ and\
  \citenamefont {Sch\"onberger}}]{cRPA_Aryasetiawan}%
  \BibitemOpen
  \bibfield  {author} {\bibinfo {author} {\bibfnamefont {F.}~\bibnamefont
  {Aryasetiawan}}, \bibinfo {author} {\bibfnamefont {K.}~\bibnamefont
  {Karlsson}}, \bibinfo {author} {\bibfnamefont {O.}~\bibnamefont {Jepsen}}, \
  and\ \bibinfo {author} {\bibfnamefont {U.}~\bibnamefont {Sch\"onberger}},\
  }\href {\doibase 10.1103/PhysRevB.74.125106} {\bibfield  {journal} {\bibinfo
  {journal} {Phys. Rev. B}\ }\textbf {\bibinfo {volume} {74}},\ \bibinfo
  {pages} {125106} (\bibinfo {year} {2006})}\BibitemShut {NoStop}%
\bibitem [{\citenamefont {Vaugier}\ \emph {et~al.}(2012)\citenamefont
  {Vaugier}, \citenamefont {Jiang},\ and\ \citenamefont
  {Biermann}}]{Vaugier_PRB2012}%
  \BibitemOpen
  \bibfield  {author} {\bibinfo {author} {\bibfnamefont {L.}~\bibnamefont
  {Vaugier}}, \bibinfo {author} {\bibfnamefont {H.}~\bibnamefont {Jiang}}, \
  and\ \bibinfo {author} {\bibfnamefont {S.}~\bibnamefont {Biermann}},\ }\href
  {\doibase 10.1103/PhysRevB.86.165105} {\bibfield  {journal} {\bibinfo
  {journal} {Phys. Rev. B}\ }\textbf {\bibinfo {volume} {86}},\ \bibinfo
  {pages} {165105} (\bibinfo {year} {2012})}\BibitemShut {NoStop}%
\bibitem [{\citenamefont {Panda}\ \emph {et~al.}(2017)\citenamefont {Panda},
  \citenamefont {Jiang},\ and\ \citenamefont {Biermann}}]{Panda_PRB}%
  \BibitemOpen
  \bibfield  {author} {\bibinfo {author} {\bibfnamefont {S.~K.}\ \bibnamefont
  {Panda}}, \bibinfo {author} {\bibfnamefont {H.}~\bibnamefont {Jiang}}, \ and\
  \bibinfo {author} {\bibfnamefont {S.}~\bibnamefont {Biermann}},\ }\href
  {\doibase 10.1103/PhysRevB.96.045137} {\bibfield  {journal} {\bibinfo
  {journal} {Phys. Rev. B}\ }\textbf {\bibinfo {volume} {96}},\ \bibinfo
  {pages} {045137} (\bibinfo {year} {2017})}\BibitemShut {NoStop}%
\bibitem [{\citenamefont {van Roekeghem}\ \emph {et~al.}(2016)\citenamefont
  {van Roekeghem}, \citenamefont {Vaugier}, \citenamefont {Jiang},\ and\
  \citenamefont {Biermann}}]{VanRoekeghem_PRB2016}%
  \BibitemOpen
  \bibfield  {author} {\bibinfo {author} {\bibfnamefont {A.}~\bibnamefont {van
  Roekeghem}}, \bibinfo {author} {\bibfnamefont {L.}~\bibnamefont {Vaugier}},
  \bibinfo {author} {\bibfnamefont {H.}~\bibnamefont {Jiang}}, \ and\ \bibinfo
  {author} {\bibfnamefont {S.}~\bibnamefont {Biermann}},\ }\href {\doibase
  10.1103/PhysRevB.94.125147} {\bibfield  {journal} {\bibinfo  {journal} {Phys.
  Rev. B}\ }\textbf {\bibinfo {volume} {94}},\ \bibinfo {pages} {125147}
  (\bibinfo {year} {2016})}\BibitemShut {NoStop}%
\bibitem [{\citenamefont {Hansmann}\ \emph {et~al.}(2013)\citenamefont
  {Hansmann}, \citenamefont {Ayral}, \citenamefont {Vaugier}, \citenamefont
  {Werner},\ and\ \citenamefont {Biermann}}]{Hansmann_PRL2013}%
  \BibitemOpen
  \bibfield  {author} {\bibinfo {author} {\bibfnamefont {P.}~\bibnamefont
  {Hansmann}}, \bibinfo {author} {\bibfnamefont {T.}~\bibnamefont {Ayral}},
  \bibinfo {author} {\bibfnamefont {L.}~\bibnamefont {Vaugier}}, \bibinfo
  {author} {\bibfnamefont {P.}~\bibnamefont {Werner}}, \ and\ \bibinfo {author}
  {\bibfnamefont {S.}~\bibnamefont {Biermann}},\ }\href {\doibase
  10.1103/PhysRevLett.110.166401} {\bibfield  {journal} {\bibinfo  {journal}
  {Phys. Rev. Lett.}\ }\textbf {\bibinfo {volume} {110}},\ \bibinfo {pages}
  {166401} (\bibinfo {year} {2013})}\BibitemShut {NoStop}%
\bibitem [{\citenamefont {Hirsch}(1985)}]{hubbard_antiferromagnet_hirsch}%
  \BibitemOpen
  \bibfield  {author} {\bibinfo {author} {\bibfnamefont {J.~E.}\ \bibnamefont
  {Hirsch}},\ }\href {\doibase 10.1103/PhysRevB.31.4403} {\bibfield  {journal}
  {\bibinfo  {journal} {Phys. Rev. B}\ }\textbf {\bibinfo {volume} {31}},\
  \bibinfo {pages} {4403} (\bibinfo {year} {1985})}\BibitemShut {NoStop}%
\bibitem [{\citenamefont {Vitali}\ \emph {et~al.}(2016)\citenamefont {Vitali},
  \citenamefont {Shi}, \citenamefont {Qin},\ and\ \citenamefont
  {Zhang}}]{hubbard_antiferromagnet_vitaly}%
  \BibitemOpen
  \bibfield  {author} {\bibinfo {author} {\bibfnamefont {E.}~\bibnamefont
  {Vitali}}, \bibinfo {author} {\bibfnamefont {H.}~\bibnamefont {Shi}},
  \bibinfo {author} {\bibfnamefont {M.}~\bibnamefont {Qin}}, \ and\ \bibinfo
  {author} {\bibfnamefont {S.}~\bibnamefont {Zhang}},\ }\href {\doibase
  10.1103/PhysRevB.94.085140} {\bibfield  {journal} {\bibinfo  {journal} {Phys.
  Rev. B}\ }\textbf {\bibinfo {volume} {94}},\ \bibinfo {pages} {085140}
  (\bibinfo {year} {2016})}\BibitemShut {NoStop}%
\bibitem [{\citenamefont {White}\ \emph {et~al.}(1989)\citenamefont {White},
  \citenamefont {Scalapino}, \citenamefont {Sugar}, \citenamefont {Loh},
  \citenamefont {Gubernatis},\ and\ \citenamefont
  {Scalettar}}]{scalapino_hubbard}%
  \BibitemOpen
  \bibfield  {author} {\bibinfo {author} {\bibfnamefont {S.~R.}\ \bibnamefont
  {White}}, \bibinfo {author} {\bibfnamefont {D.~J.}\ \bibnamefont
  {Scalapino}}, \bibinfo {author} {\bibfnamefont {R.~L.}\ \bibnamefont
  {Sugar}}, \bibinfo {author} {\bibfnamefont {E.~Y.}\ \bibnamefont {Loh}},
  \bibinfo {author} {\bibfnamefont {J.~E.}\ \bibnamefont {Gubernatis}}, \ and\
  \bibinfo {author} {\bibfnamefont {R.~T.}\ \bibnamefont {Scalettar}},\ }\href
  {\doibase 10.1103/PhysRevB.40.506} {\bibfield  {journal} {\bibinfo  {journal}
  {Phys. Rev. B}\ }\textbf {\bibinfo {volume} {40}},\ \bibinfo {pages} {506}
  (\bibinfo {year} {1989})}\BibitemShut {NoStop}%
\bibitem [{\citenamefont {Corboz}\ \emph {et~al.}(2014)\citenamefont {Corboz},
  \citenamefont {Rice},\ and\ \citenamefont {Troyer}}]{tJ_rice_troyer}%
  \BibitemOpen
  \bibfield  {author} {\bibinfo {author} {\bibfnamefont {P.}~\bibnamefont
  {Corboz}}, \bibinfo {author} {\bibfnamefont {T.~M.}\ \bibnamefont {Rice}}, \
  and\ \bibinfo {author} {\bibfnamefont {M.}~\bibnamefont {Troyer}},\ }\href
  {\doibase 10.1103/PhysRevLett.113.046402} {\bibfield  {journal} {\bibinfo
  {journal} {Phys. Rev. Lett.}\ }\textbf {\bibinfo {volume} {113}},\ \bibinfo
  {pages} {046402} (\bibinfo {year} {2014})}\BibitemShut {NoStop}%
\bibitem [{\citenamefont {Qin}\ \emph {et~al.}(2020)\citenamefont {Qin},
  \citenamefont {Chung}, \citenamefont {Shi}, \citenamefont {Vitali},
  \citenamefont {Hubig}, \citenamefont {Schollw\"ock}, \citenamefont {White},\
  and\ \citenamefont {Zhang}}]{hubbard_absence_of_SC}%
  \BibitemOpen
  \bibfield  {author} {\bibinfo {author} {\bibfnamefont {M.}~\bibnamefont
  {Qin}}, \bibinfo {author} {\bibfnamefont {C.-M.}\ \bibnamefont {Chung}},
  \bibinfo {author} {\bibfnamefont {H.}~\bibnamefont {Shi}}, \bibinfo {author}
  {\bibfnamefont {E.}~\bibnamefont {Vitali}}, \bibinfo {author} {\bibfnamefont
  {C.}~\bibnamefont {Hubig}}, \bibinfo {author} {\bibfnamefont
  {U.}~\bibnamefont {Schollw\"ock}}, \bibinfo {author} {\bibfnamefont {S.~R.}\
  \bibnamefont {White}}, \ and\ \bibinfo {author} {\bibfnamefont
  {S.}~\bibnamefont {Zhang}} (\bibinfo {collaboration} {Simons Collaboration on
  the Many-Electron Problem}),\ }\href {\doibase 10.1103/PhysRevX.10.031016}
  {\bibfield  {journal} {\bibinfo  {journal} {Phys. Rev. X}\ }\textbf {\bibinfo
  {volume} {10}},\ \bibinfo {pages} {031016} (\bibinfo {year}
  {2020})}\BibitemShut {NoStop}%
\bibitem [{\citenamefont {Werner}\ \emph {et~al.}(2016)\citenamefont {Werner},
  \citenamefont {Hoshino},\ and\ \citenamefont
  {Shinaoka}}]{spin_freezing_cuprates}%
  \BibitemOpen
  \bibfield  {author} {\bibinfo {author} {\bibfnamefont {P.}~\bibnamefont
  {Werner}}, \bibinfo {author} {\bibfnamefont {S.}~\bibnamefont {Hoshino}}, \
  and\ \bibinfo {author} {\bibfnamefont {H.}~\bibnamefont {Shinaoka}},\ }\href
  {\doibase 10.1103/PhysRevB.94.245134} {\bibfield  {journal} {\bibinfo
  {journal} {Phys. Rev. B}\ }\textbf {\bibinfo {volume} {94}},\ \bibinfo
  {pages} {245134} (\bibinfo {year} {2016})}\BibitemShut {NoStop}%
\bibitem [{\citenamefont {Martins}\ \emph {et~al.}(2011)\citenamefont
  {Martins}, \citenamefont {Aichhorn}, \citenamefont {Vaugier},\ and\
  \citenamefont {Biermann}}]{Martins1}%
  \BibitemOpen
  \bibfield  {author} {\bibinfo {author} {\bibfnamefont {C.}~\bibnamefont
  {Martins}}, \bibinfo {author} {\bibfnamefont {M.}~\bibnamefont {Aichhorn}},
  \bibinfo {author} {\bibfnamefont {L.}~\bibnamefont {Vaugier}}, \ and\
  \bibinfo {author} {\bibfnamefont {S.}~\bibnamefont {Biermann}},\ }\href
  {\doibase 10.1103/PhysRevLett.107.266404} {\bibfield  {journal} {\bibinfo
  {journal} {Phys. Rev. Lett.}\ }\textbf {\bibinfo {volume} {107}},\ \bibinfo
  {pages} {266404} (\bibinfo {year} {2011})}\BibitemShut {NoStop}%
\bibitem [{\citenamefont {Martins}\ \emph {et~al.}(2017)\citenamefont
  {Martins}, \citenamefont {Aichhorn},\ and\ \citenamefont
  {Biermann}}]{Martins_2017}%
  \BibitemOpen
  \bibfield  {author} {\bibinfo {author} {\bibfnamefont {C.}~\bibnamefont
  {Martins}}, \bibinfo {author} {\bibfnamefont {M.}~\bibnamefont {Aichhorn}}, \
  and\ \bibinfo {author} {\bibfnamefont {S.}~\bibnamefont {Biermann}},\ }\href
  {\doibase 10.1088/1361-648x/aa648f} {\bibfield  {journal} {\bibinfo
  {journal} {Journal of Physics: Condensed Matter}\ }\textbf {\bibinfo {volume}
  {29}},\ \bibinfo {pages} {263001} (\bibinfo {year} {2017})}\BibitemShut
  {NoStop}%
\bibitem [{\citenamefont {Poteryaev}\ \emph {et~al.}(2007)\citenamefont
  {Poteryaev}, \citenamefont {Tomczak}, \citenamefont {Biermann}, \citenamefont
  {Georges}, \citenamefont {Lichtenstein}, \citenamefont {Rubtsov},
  \citenamefont {Saha-Dasgupta},\ and\ \citenamefont {Andersen}}]{Poteryaev}%
  \BibitemOpen
  \bibfield  {author} {\bibinfo {author} {\bibfnamefont {A.~I.}\ \bibnamefont
  {Poteryaev}}, \bibinfo {author} {\bibfnamefont {J.~M.}\ \bibnamefont
  {Tomczak}}, \bibinfo {author} {\bibfnamefont {S.}~\bibnamefont {Biermann}},
  \bibinfo {author} {\bibfnamefont {A.}~\bibnamefont {Georges}}, \bibinfo
  {author} {\bibfnamefont {A.~I.}\ \bibnamefont {Lichtenstein}}, \bibinfo
  {author} {\bibfnamefont {A.~N.}\ \bibnamefont {Rubtsov}}, \bibinfo {author}
  {\bibfnamefont {T.}~\bibnamefont {Saha-Dasgupta}}, \ and\ \bibinfo {author}
  {\bibfnamefont {O.~K.}\ \bibnamefont {Andersen}},\ }\href {\doibase
  10.1103/PhysRevB.76.085127} {\bibfield  {journal} {\bibinfo  {journal} {Phys.
  Rev. B}\ }\textbf {\bibinfo {volume} {76}},\ \bibinfo {pages} {085127}
  (\bibinfo {year} {2007})}\BibitemShut {NoStop}%
\bibitem [{\citenamefont {Hirayama}\ \emph {et~al.}(2017)\citenamefont
  {Hirayama}, \citenamefont {Miyake}, \citenamefont {Imada},\ and\
  \citenamefont {Biermann}}]{Hirayama}%
  \BibitemOpen
  \bibfield  {author} {\bibinfo {author} {\bibfnamefont {M.}~\bibnamefont
  {Hirayama}}, \bibinfo {author} {\bibfnamefont {T.}~\bibnamefont {Miyake}},
  \bibinfo {author} {\bibfnamefont {M.}~\bibnamefont {Imada}}, \ and\ \bibinfo
  {author} {\bibfnamefont {S.}~\bibnamefont {Biermann}},\ }\href {\doibase
  10.1103/PhysRevB.96.075102} {\bibfield  {journal} {\bibinfo  {journal} {Phys.
  Rev. B}\ }\textbf {\bibinfo {volume} {96}},\ \bibinfo {pages} {075102}
  (\bibinfo {year} {2017})}\BibitemShut {NoStop}%
\bibitem [{\citenamefont {Brandow}(1977)}]{brandow_dp}%
  \BibitemOpen
  \bibfield  {author} {\bibinfo {author} {\bibfnamefont {B.}~\bibnamefont
  {Brandow}},\ }\href {\doibase 10.1080/00018737700101443} {\bibfield
  {journal} {\bibinfo  {journal} {Advances in Physics}\ }\textbf {\bibinfo
  {volume} {26}},\ \bibinfo {pages} {651} (\bibinfo {year} {1977})},\ \Eprint
  {http://arxiv.org/abs/https://doi.org/10.1080/00018737700101443}
  {https://doi.org/10.1080/00018737700101443} \BibitemShut {NoStop}%
\bibitem [{\citenamefont {Herring}(1966)}]{perfect_screening_herring}%
  \BibitemOpen
  \bibfield  {author} {\bibinfo {author} {\bibfnamefont {C.}~\bibnamefont
  {Herring}},\ }\href@noop {} {\emph {\bibinfo {title} {Magnetism: Exchange
  Interactions Among Itinerant Electrons}}},\ Vol.~\bibinfo {volume} {4}\
  (\bibinfo  {publisher} {Academic Press, New York},\ \bibinfo {year}
  {1966})\BibitemShut {NoStop}%
\bibitem [{\citenamefont {Fujimori}\ and\ \citenamefont
  {Minami}(1984)}]{FujimoriMinami}%
  \BibitemOpen
  \bibfield  {author} {\bibinfo {author} {\bibfnamefont {A.}~\bibnamefont
  {Fujimori}}\ and\ \bibinfo {author} {\bibfnamefont {F.}~\bibnamefont
  {Minami}},\ }\href {\doibase 10.1103/PhysRevB.30.957} {\bibfield  {journal}
  {\bibinfo  {journal} {Phys. Rev. B}\ }\textbf {\bibinfo {volume} {30}},\
  \bibinfo {pages} {957} (\bibinfo {year} {1984})}\BibitemShut {NoStop}%
\bibitem [{\citenamefont {Seth}\ \emph {et~al.}(2017)\citenamefont {Seth},
  \citenamefont {Hansmann}, \citenamefont {van Roekeghem}, \citenamefont
  {Vaugier},\ and\ \citenamefont {Biermann}}]{shell_folding}%
  \BibitemOpen
  \bibfield  {author} {\bibinfo {author} {\bibfnamefont {P.}~\bibnamefont
  {Seth}}, \bibinfo {author} {\bibfnamefont {P.}~\bibnamefont {Hansmann}},
  \bibinfo {author} {\bibfnamefont {A.}~\bibnamefont {van Roekeghem}}, \bibinfo
  {author} {\bibfnamefont {L.}~\bibnamefont {Vaugier}}, \ and\ \bibinfo
  {author} {\bibfnamefont {S.}~\bibnamefont {Biermann}},\ }\href {\doibase
  10.1103/PhysRevLett.119.056401} {\bibfield  {journal} {\bibinfo  {journal}
  {Phys. Rev. Lett.}\ }\textbf {\bibinfo {volume} {119}},\ \bibinfo {pages}
  {056401} (\bibinfo {year} {2017})}\BibitemShut {NoStop}%
\bibitem [{\citenamefont {Han}\ \emph {et~al.}(2011)\citenamefont {Han},
  \citenamefont {Wang}, \citenamefont {Marianetti},\ and\ \citenamefont
  {Millis}}]{dp_millis}%
  \BibitemOpen
  \bibfield  {author} {\bibinfo {author} {\bibfnamefont {M.~J.}\ \bibnamefont
  {Han}}, \bibinfo {author} {\bibfnamefont {X.}~\bibnamefont {Wang}}, \bibinfo
  {author} {\bibfnamefont {C.~A.}\ \bibnamefont {Marianetti}}, \ and\ \bibinfo
  {author} {\bibfnamefont {A.~J.}\ \bibnamefont {Millis}},\ }\href {\doibase
  10.1103/PhysRevLett.107.206804} {\bibfield  {journal} {\bibinfo  {journal}
  {Phys. Rev. Lett.}\ }\textbf {\bibinfo {volume} {107}},\ \bibinfo {pages}
  {206804} (\bibinfo {year} {2011})}\BibitemShut {NoStop}%
\bibitem [{\citenamefont {Parragh}\ \emph {et~al.}(2013)\citenamefont
  {Parragh}, \citenamefont {Sangiovanni}, \citenamefont {Hansmann},
  \citenamefont {Hummel}, \citenamefont {Held},\ and\ \citenamefont
  {Toschi}}]{d_dp_parragh}%
  \BibitemOpen
  \bibfield  {author} {\bibinfo {author} {\bibfnamefont {N.}~\bibnamefont
  {Parragh}}, \bibinfo {author} {\bibfnamefont {G.}~\bibnamefont
  {Sangiovanni}}, \bibinfo {author} {\bibfnamefont {P.}~\bibnamefont
  {Hansmann}}, \bibinfo {author} {\bibfnamefont {S.}~\bibnamefont {Hummel}},
  \bibinfo {author} {\bibfnamefont {K.}~\bibnamefont {Held}}, \ and\ \bibinfo
  {author} {\bibfnamefont {A.}~\bibnamefont {Toschi}},\ }\href {\doibase
  10.1103/PhysRevB.88.195116} {\bibfield  {journal} {\bibinfo  {journal} {Phys.
  Rev. B}\ }\textbf {\bibinfo {volume} {88}},\ \bibinfo {pages} {195116}
  (\bibinfo {year} {2013})}\BibitemShut {NoStop}%
\bibitem [{\citenamefont {Zaanen}\ \emph {et~al.}(1985)\citenamefont {Zaanen},
  \citenamefont {Sawatzky},\ and\ \citenamefont
  {Allen}}]{ZaanenSawatzkyAllenPRL}%
  \BibitemOpen
  \bibfield  {author} {\bibinfo {author} {\bibfnamefont {J.}~\bibnamefont
  {Zaanen}}, \bibinfo {author} {\bibfnamefont {G.~A.}\ \bibnamefont
  {Sawatzky}}, \ and\ \bibinfo {author} {\bibfnamefont {J.~W.}\ \bibnamefont
  {Allen}},\ }\href {\doibase 10.1103/PhysRevLett.55.418} {\bibfield  {journal}
  {\bibinfo  {journal} {Phys. Rev. Lett.}\ }\textbf {\bibinfo {volume} {55}},\
  \bibinfo {pages} {418} (\bibinfo {year} {1985})}\BibitemShut {NoStop}%
\bibitem [{\citenamefont {Sawatzky}\ and\ \citenamefont
  {Allen}(1984)}]{Sawatzky_NIO_cluster}%
  \BibitemOpen
  \bibfield  {author} {\bibinfo {author} {\bibfnamefont {G.~A.}\ \bibnamefont
  {Sawatzky}}\ and\ \bibinfo {author} {\bibfnamefont {J.~W.}\ \bibnamefont
  {Allen}},\ }\href {\doibase 10.1103/PhysRevLett.53.2339} {\bibfield
  {journal} {\bibinfo  {journal} {Phys. Rev. Lett.}\ }\textbf {\bibinfo
  {volume} {53}},\ \bibinfo {pages} {2339} (\bibinfo {year}
  {1984})}\BibitemShut {NoStop}%
\bibitem [{\citenamefont {Florens}\ and\ \citenamefont
  {Georges}(2002)}]{SR_impurity}%
  \BibitemOpen
  \bibfield  {author} {\bibinfo {author} {\bibfnamefont {S.}~\bibnamefont
  {Florens}}\ and\ \bibinfo {author} {\bibfnamefont {A.}~\bibnamefont
  {Georges}},\ }\href {\doibase 10.1103/PhysRevB.66.165111} {\bibfield
  {journal} {\bibinfo  {journal} {Phys. Rev. B}\ }\textbf {\bibinfo {volume}
  {66}},\ \bibinfo {pages} {165111} (\bibinfo {year} {2002})}\BibitemShut
  {NoStop}%
\bibitem [{\citenamefont {Florens}\ and\ \citenamefont
  {Georges}(2004)}]{SR_meanfield}%
  \BibitemOpen
  \bibfield  {author} {\bibinfo {author} {\bibfnamefont {S.}~\bibnamefont
  {Florens}}\ and\ \bibinfo {author} {\bibfnamefont {A.}~\bibnamefont
  {Georges}},\ }\href {\doibase 10.1103/PhysRevB.70.035114} {\bibfield
  {journal} {\bibinfo  {journal} {Phys. Rev. B}\ }\textbf {\bibinfo {volume}
  {70}},\ \bibinfo {pages} {035114} (\bibinfo {year} {2004})}\BibitemShut
  {NoStop}%
\bibitem [{\citenamefont {Krivenko}\ and\ \citenamefont
  {Biermann}(2015)}]{SR_Krivenko}%
  \BibitemOpen
  \bibfield  {author} {\bibinfo {author} {\bibfnamefont {I.~S.}\ \bibnamefont
  {Krivenko}}\ and\ \bibinfo {author} {\bibfnamefont {S.}~\bibnamefont
  {Biermann}},\ }\href {\doibase 10.1103/PhysRevB.91.155149} {\bibfield
  {journal} {\bibinfo  {journal} {Phys. Rev. B}\ }\textbf {\bibinfo {volume}
  {91}},\ \bibinfo {pages} {155149} (\bibinfo {year} {2015})}\BibitemShut
  {NoStop}%
\bibitem [{\citenamefont {Zhang}\ and\ \citenamefont {Rice}(1988)}]{ZhangRice}%
  \BibitemOpen
  \bibfield  {author} {\bibinfo {author} {\bibfnamefont {F.~C.}\ \bibnamefont
  {Zhang}}\ and\ \bibinfo {author} {\bibfnamefont {T.~M.}\ \bibnamefont
  {Rice}},\ }\href {\doibase 10.1103/PhysRevB.37.3759} {\bibfield  {journal}
  {\bibinfo  {journal} {Phys. Rev. B}\ }\textbf {\bibinfo {volume} {37}},\
  \bibinfo {pages} {3759} (\bibinfo {year} {1988})}\BibitemShut {NoStop}%
\bibitem [{\citenamefont {Steinbauer}(2019{\natexlab{a}})}]{MyThesis}%
  \BibitemOpen
  \bibfield  {author} {\bibinfo {author} {\bibfnamefont {J.}~\bibnamefont
  {Steinbauer}},\ }\emph {\bibinfo {title} {Multi-Orbital Physics in Materials
  with Strong Electronic Correlations: Hund's Coupling and Inter-Shell
  Interactions}},\ \href@noop {} {Ph.D. thesis},\ \bibinfo  {school} {Th\`ese
  de doctorat de l'Universit\'e Paris-Saclay pr\'epar\'ee \`a \'Ecole
  Polytechnique} (\bibinfo {year} {2019}{\natexlab{a}})\BibitemShut {NoStop}%
\bibitem [{\citenamefont {Hansmann}\ \emph {et~al.}(2014)\citenamefont
  {Hansmann}, \citenamefont {Parragh}, \citenamefont {Toschi}, \citenamefont
  {Sangiovanni},\ and\ \citenamefont {Held}}]{Hansmann_2014}%
  \BibitemOpen
  \bibfield  {author} {\bibinfo {author} {\bibfnamefont {P.}~\bibnamefont
  {Hansmann}}, \bibinfo {author} {\bibfnamefont {N.}~\bibnamefont {Parragh}},
  \bibinfo {author} {\bibfnamefont {A.}~\bibnamefont {Toschi}}, \bibinfo
  {author} {\bibfnamefont {G.}~\bibnamefont {Sangiovanni}}, \ and\ \bibinfo
  {author} {\bibfnamefont {K.}~\bibnamefont {Held}},\ }\href {\doibase
  10.1088/1367-2630/16/3/033009} {\bibfield  {journal} {\bibinfo  {journal}
  {New Journal of Physics}\ }\textbf {\bibinfo {volume} {16}},\ \bibinfo
  {pages} {033009} (\bibinfo {year} {2014})}\BibitemShut {NoStop}%
\bibitem [{\citenamefont {Georges}\ \emph {et~al.}(1996)\citenamefont
  {Georges}, \citenamefont {Kotliar}, \citenamefont {Krauth},\ and\
  \citenamefont {Rozenberg}}]{DMFTkotliar_georges}%
  \BibitemOpen
  \bibfield  {author} {\bibinfo {author} {\bibfnamefont {A.}~\bibnamefont
  {Georges}}, \bibinfo {author} {\bibfnamefont {G.}~\bibnamefont {Kotliar}},
  \bibinfo {author} {\bibfnamefont {W.}~\bibnamefont {Krauth}}, \ and\ \bibinfo
  {author} {\bibfnamefont {M.~J.}\ \bibnamefont {Rozenberg}},\ }\href {\doibase
  10.1103/RevModPhys.68.13} {\bibfield  {journal} {\bibinfo  {journal} {Rev.
  Mod. Phys.}\ }\textbf {\bibinfo {volume} {68}},\ \bibinfo {pages} {13}
  (\bibinfo {year} {1996})}\BibitemShut {NoStop}%
\bibitem [{\citenamefont {Shinaoka}\ \emph {et~al.}(2015)\citenamefont
  {Shinaoka}, \citenamefont {Nomura}, \citenamefont {Biermann}, \citenamefont
  {Troyer},\ and\ \citenamefont {Werner}}]{QMC_negativesign}%
  \BibitemOpen
  \bibfield  {author} {\bibinfo {author} {\bibfnamefont {H.}~\bibnamefont
  {Shinaoka}}, \bibinfo {author} {\bibfnamefont {Y.}~\bibnamefont {Nomura}},
  \bibinfo {author} {\bibfnamefont {S.}~\bibnamefont {Biermann}}, \bibinfo
  {author} {\bibfnamefont {M.}~\bibnamefont {Troyer}}, \ and\ \bibinfo {author}
  {\bibfnamefont {P.}~\bibnamefont {Werner}},\ }\href {\doibase
  10.1103/PhysRevB.92.195126} {\bibfield  {journal} {\bibinfo  {journal} {Phys.
  Rev. B}\ }\textbf {\bibinfo {volume} {92}},\ \bibinfo {pages} {195126}
  (\bibinfo {year} {2015})}\BibitemShut {NoStop}%
\bibitem [{Note1()}]{Note1}%
  \BibitemOpen
  \bibinfo {note} {Within the Hartree-Fock approximation, the p-p and
  intershell interactions are approximated on a mean-field level. Neglecting
  any spin-flipping terms, we get \begin {align*}\label {eq:Hartree_Fock_dp}
  \begin {split} U^{pp} n^{p}_{\delimiter "3222378 }n^{p}_{\delimiter "3223379
  } &+ U^{dp} (n^{d}_{\delimiter "3222378 } + n^{d}_{\delimiter "3223379
  })(n^{p}_{\delimiter "3222378 } + n^{p}_{\delimiter "3223379 })\\ & \approx
  U^{pp} \left \protect \{ n^{p}_{\delimiter "3222378 }\protect \Braket
  {n^{p}_{\delimiter "3223379 }} + \protect \Braket {n^{p}_{\delimiter "3222378
  }}n^{p}_{\delimiter "3223379 } + \protect \Braket {n^{p}_{\delimiter "3222378
  }}\protect \Braket {n^{p}_{\delimiter "3223379 }} \right \protect \}\\
  &+U^{dp} \left \protect \{(n^{d}_{\delimiter "3222378 } + n^{d}_{\delimiter
  "3223379 }) \protect \Braket {n^{p}_{\delimiter "3222378 } +
  n^{p}_{\delimiter "3223379 }} + \protect \Braket {n^{d}_{\delimiter "3222378
  } + n^{d}_{\delimiter "3223379 }}(n^{p}_{\delimiter "3222378 }+
  n^{p}_{\delimiter "3223379 })\right .\\ &\left . + \protect \Braket
  {n^{d}_{\delimiter "3222378 } + n^{d}_{\delimiter "3223379 }}\protect \Braket
  {n^{p}_{\delimiter "3222378 } + n^{p}_{\delimiter "3223379 }}\right \protect
  \}\\ &-U^{dp}\left \protect \{ d^{\dagger }_{\delimiter "3222378
  }p_{\delimiter "3222378 }\protect \Braket {p^{\dagger }_{\delimiter "3222378
  }d_{\delimiter "3222378 }} + \protect \Braket {d^{\dagger }_{\delimiter
  "3222378 }p_{\delimiter "3222378 }}p^{\dagger }_{\delimiter "3222378
  }d_{\delimiter "3222378 } + \protect \Braket {d^{\dagger }_{\delimiter
  "3222378 }p_{\delimiter "3222378 }}\protect \Braket {p^{\dagger }_{\delimiter
  "3222378 }d_{\delimiter "3222378 }}\right \protect \}\\ &-U^{dp}\left
  \protect \{ d^{\dagger }_{\delimiter "3223379 }p_{\delimiter "3223379
  }\protect \Braket {p^{\dagger }_{\delimiter "3223379 }d_{\delimiter "3223379
  }} + \protect \Braket {d^{\dagger }_{\delimiter "3223379 }p_{\delimiter
  "3223379 }}p^{\dagger }_{\delimiter "3223379 }d_{\delimiter "3223379 } +
  \protect \Braket {d^{\dagger }_{\delimiter "3223379 }p_{\delimiter "3223379
  }}\protect \Braket {p^{\dagger }_{\delimiter "3223379 }d_{\delimiter "3223379
  }}\right \protect \} \protect \text { .} \end {split} \end {align*} Within
  this approximation, the p electrons are thus treated as being effectively
  uncorrelated.}\BibitemShut {Stop}%
\bibitem [{\citenamefont {Steinbauer}(2019{\natexlab{b}})}]{Jakob_thesis}%
  \BibitemOpen
  \bibfield  {author} {\bibinfo {author} {\bibfnamefont {J.}~\bibnamefont
  {Steinbauer}},\ }\emph {\bibinfo {title} {Multi-Orbital Physics in Materials
  with Strong Electronic Correlations: Hund's Coupling and Inter-Shell
  Interactions}},\ \href@noop {} {Ph.D. thesis},\ \bibinfo  {school} {\'Ecole
  Polytechnique} (\bibinfo {year} {2019}{\natexlab{b}})\BibitemShut {NoStop}%
\bibitem [{\citenamefont {Georgescu}\ and\ \citenamefont
  {Ismail-Beigi}(2015)}]{Generalized_SP}%
  \BibitemOpen
  \bibfield  {author} {\bibinfo {author} {\bibfnamefont {A.~B.}\ \bibnamefont
  {Georgescu}}\ and\ \bibinfo {author} {\bibfnamefont {S.}~\bibnamefont
  {Ismail-Beigi}},\ }\href {\doibase 10.1103/PhysRevB.92.235117} {\bibfield
  {journal} {\bibinfo  {journal} {Phys. Rev. B}\ }\textbf {\bibinfo {volume}
  {92}},\ \bibinfo {pages} {235117} (\bibinfo {year} {2015})}\BibitemShut
  {NoStop}%
\bibitem [{\citenamefont {Levy}\ \emph {et~al.}(2017)\citenamefont {Levy},
  \citenamefont {LeBlanc},\ and\ \citenamefont {Gull}}]{maxent_Gull}%
  \BibitemOpen
  \bibfield  {author} {\bibinfo {author} {\bibfnamefont {R.}~\bibnamefont
  {Levy}}, \bibinfo {author} {\bibfnamefont {J.}~\bibnamefont {LeBlanc}}, \
  and\ \bibinfo {author} {\bibfnamefont {E.}~\bibnamefont {Gull}},\ }\href
  {\doibase https://doi.org/10.1016/j.cpc.2017.01.018} {\bibfield  {journal}
  {\bibinfo  {journal} {Computer Physics Communications}\ }\textbf {\bibinfo
  {volume} {215}},\ \bibinfo {pages} {149 } (\bibinfo {year}
  {2017})}\BibitemShut {NoStop}%
\bibitem [{\citenamefont {Ayral}\ \emph {et~al.}(2012)\citenamefont {Ayral},
  \citenamefont {Werner},\ and\ \citenamefont {Biermann}}]{ayral_spectral}%
  \BibitemOpen
  \bibfield  {author} {\bibinfo {author} {\bibfnamefont {T.}~\bibnamefont
  {Ayral}}, \bibinfo {author} {\bibfnamefont {P.}~\bibnamefont {Werner}}, \
  and\ \bibinfo {author} {\bibfnamefont {S.}~\bibnamefont {Biermann}},\ }\href
  {\doibase 10.1103/PhysRevLett.109.226401} {\bibfield  {journal} {\bibinfo
  {journal} {Phys. Rev. Lett.}\ }\textbf {\bibinfo {volume} {109}},\ \bibinfo
  {pages} {226401} (\bibinfo {year} {2012})}\BibitemShut {NoStop}%
\bibitem [{\citenamefont {Emery}(1987)}]{Emery_model}%
  \BibitemOpen
  \bibfield  {author} {\bibinfo {author} {\bibfnamefont {V.~J.}\ \bibnamefont
  {Emery}},\ }\href {\doibase 10.1103/PhysRevLett.58.2794} {\bibfield
  {journal} {\bibinfo  {journal} {Phys. Rev. Lett.}\ }\textbf {\bibinfo
  {volume} {58}},\ \bibinfo {pages} {2794} (\bibinfo {year}
  {1987})}\BibitemShut {NoStop}%
\bibitem [{\citenamefont {Andersen}\ \emph {et~al.}(1995)\citenamefont
  {Andersen}, \citenamefont {Liechtenstein}, \citenamefont {Jepsen},\ and\
  \citenamefont {Paulsen}}]{Andersen_Emery}%
  \BibitemOpen
  \bibfield  {author} {\bibinfo {author} {\bibfnamefont {O.}~\bibnamefont
  {Andersen}}, \bibinfo {author} {\bibfnamefont {A.}~\bibnamefont
  {Liechtenstein}}, \bibinfo {author} {\bibfnamefont {O.}~\bibnamefont
  {Jepsen}}, \ and\ \bibinfo {author} {\bibfnamefont {F.}~\bibnamefont
  {Paulsen}},\ }\href {\doibase https://doi.org/10.1016/0022-3697(95)00269-3}
  {\bibfield  {journal} {\bibinfo  {journal} {Journal of Physics and Chemistry
  of Solids}\ }\textbf {\bibinfo {volume} {56}},\ \bibinfo {pages} {1573 }
  (\bibinfo {year} {1995})},\ \bibinfo {note} {proceedings of the Conference on
  Spectroscopies in Novel Superconductors}\BibitemShut {NoStop}%
\bibitem [{\citenamefont {Karolak}\ \emph {et~al.}(2010)\citenamefont
  {Karolak}, \citenamefont {Ulm}, \citenamefont {Wehling}, \citenamefont
  {Mazurenko}, \citenamefont {Poteryaev},\ and\ \citenamefont
  {Lichtenstein}}]{NiO_dc}%
  \BibitemOpen
  \bibfield  {author} {\bibinfo {author} {\bibfnamefont {M.}~\bibnamefont
  {Karolak}}, \bibinfo {author} {\bibfnamefont {G.}~\bibnamefont {Ulm}},
  \bibinfo {author} {\bibfnamefont {T.}~\bibnamefont {Wehling}}, \bibinfo
  {author} {\bibfnamefont {V.}~\bibnamefont {Mazurenko}}, \bibinfo {author}
  {\bibfnamefont {A.}~\bibnamefont {Poteryaev}}, \ and\ \bibinfo {author}
  {\bibfnamefont {A.}~\bibnamefont {Lichtenstein}},\ }\href {\doibase
  https://doi.org/10.1016/j.elspec.2010.05.021} {\bibfield  {journal} {\bibinfo
   {journal} {Journal of Electron Spectroscopy and Related Phenomena}\ }\textbf
  {\bibinfo {volume} {181}},\ \bibinfo {pages} {11 } (\bibinfo {year}
  {2010})},\ \bibinfo {note} {proceedings of International Workshop on Strong
  Correlations and Angle-Resolved Photoemission Spectroscopy 2009}\BibitemShut
  {NoStop}%
\bibitem [{\citenamefont {Wang}\ \emph {et~al.}(2012)\citenamefont {Wang},
  \citenamefont {Han}, \citenamefont {de' Medici}, \citenamefont {Park},
  \citenamefont {Marianetti},\ and\ \citenamefont {Millis}}]{dp_millis_medici}%
  \BibitemOpen
  \bibfield  {author} {\bibinfo {author} {\bibfnamefont {X.}~\bibnamefont
  {Wang}}, \bibinfo {author} {\bibfnamefont {M.~J.}\ \bibnamefont {Han}},
  \bibinfo {author} {\bibfnamefont {L.}~\bibnamefont {de' Medici}}, \bibinfo
  {author} {\bibfnamefont {H.}~\bibnamefont {Park}}, \bibinfo {author}
  {\bibfnamefont {C.~A.}\ \bibnamefont {Marianetti}}, \ and\ \bibinfo {author}
  {\bibfnamefont {A.~J.}\ \bibnamefont {Millis}},\ }\href {\doibase
  10.1103/PhysRevB.86.195136} {\bibfield  {journal} {\bibinfo  {journal} {Phys.
  Rev. B}\ }\textbf {\bibinfo {volume} {86}},\ \bibinfo {pages} {195136}
  (\bibinfo {year} {2012})}\BibitemShut {NoStop}%
\bibitem [{\citenamefont {Hansmann}\ \emph {et~al.}(2009)\citenamefont
  {Hansmann}, \citenamefont {Yang}, \citenamefont {Toschi}, \citenamefont
  {Khaliullin}, \citenamefont {Andersen},\ and\ \citenamefont
  {Held}}]{d_hansmann}%
  \BibitemOpen
  \bibfield  {author} {\bibinfo {author} {\bibfnamefont {P.}~\bibnamefont
  {Hansmann}}, \bibinfo {author} {\bibfnamefont {X.}~\bibnamefont {Yang}},
  \bibinfo {author} {\bibfnamefont {A.}~\bibnamefont {Toschi}}, \bibinfo
  {author} {\bibfnamefont {G.}~\bibnamefont {Khaliullin}}, \bibinfo {author}
  {\bibfnamefont {O.~K.}\ \bibnamefont {Andersen}}, \ and\ \bibinfo {author}
  {\bibfnamefont {K.}~\bibnamefont {Held}},\ }\href {\doibase
  10.1103/PhysRevLett.103.016401} {\bibfield  {journal} {\bibinfo  {journal}
  {Phys. Rev. Lett.}\ }\textbf {\bibinfo {volume} {103}},\ \bibinfo {pages}
  {016401} (\bibinfo {year} {2009})}\BibitemShut {NoStop}%
\bibitem [{\citenamefont {Wilson}(1975)}]{Wilson_NRG}%
  \BibitemOpen
  \bibfield  {author} {\bibinfo {author} {\bibfnamefont {K.~G.}\ \bibnamefont
  {Wilson}},\ }\href {\doibase 10.1103/RevModPhys.47.773} {\bibfield  {journal}
  {\bibinfo  {journal} {Rev. Mod. Phys.}\ }\textbf {\bibinfo {volume} {47}},\
  \bibinfo {pages} {773} (\bibinfo {year} {1975})}\BibitemShut {NoStop}%
\bibitem [{\citenamefont {Casula}\ \emph {et~al.}(2012)\citenamefont {Casula},
  \citenamefont {Rubtsov},\ and\ \citenamefont {Biermann}}]{DALA_Casula}%
  \BibitemOpen
  \bibfield  {author} {\bibinfo {author} {\bibfnamefont {M.}~\bibnamefont
  {Casula}}, \bibinfo {author} {\bibfnamefont {A.}~\bibnamefont {Rubtsov}}, \
  and\ \bibinfo {author} {\bibfnamefont {S.}~\bibnamefont {Biermann}},\ }\href
  {\doibase 10.1103/PhysRevB.85.035115} {\bibfield  {journal} {\bibinfo
  {journal} {Phys. Rev. B}\ }\textbf {\bibinfo {volume} {85}},\ \bibinfo
  {pages} {035115} (\bibinfo {year} {2012})}\BibitemShut {NoStop}%
\bibitem [{\citenamefont {Scalapino}(1995)}]{SCALAPINO1995329}%
  \BibitemOpen
  \bibfield  {author} {\bibinfo {author} {\bibfnamefont {D.}~\bibnamefont
  {Scalapino}},\ }\href {\doibase https://doi.org/10.1016/0370-1573(94)00086-I}
  {\bibfield  {journal} {\bibinfo  {journal} {Physics Reports}\ }\textbf
  {\bibinfo {volume} {250}},\ \bibinfo {pages} {329 } (\bibinfo {year}
  {1995})}\BibitemShut {NoStop}%
\bibitem [{\citenamefont {Miyake}\ and\ \citenamefont
  {Aryasetiawan}(2008)}]{cRPA_Miyake}%
  \BibitemOpen
  \bibfield  {author} {\bibinfo {author} {\bibfnamefont {T.}~\bibnamefont
  {Miyake}}\ and\ \bibinfo {author} {\bibfnamefont {F.}~\bibnamefont
  {Aryasetiawan}},\ }\href {\doibase 10.1103/PhysRevB.77.085122} {\bibfield
  {journal} {\bibinfo  {journal} {Phys. Rev. B}\ }\textbf {\bibinfo {volume}
  {77}},\ \bibinfo {pages} {085122} (\bibinfo {year} {2008})}\BibitemShut
  {NoStop}%
\bibitem [{\citenamefont {Nakamura}\ \emph {et~al.}(2008)\citenamefont
  {Nakamura}, \citenamefont {Arita},\ and\ \citenamefont {Imada}}]{cRPA_Imada}%
  \BibitemOpen
  \bibfield  {author} {\bibinfo {author} {\bibfnamefont {K.}~\bibnamefont
  {Nakamura}}, \bibinfo {author} {\bibfnamefont {R.}~\bibnamefont {Arita}}, \
  and\ \bibinfo {author} {\bibfnamefont {M.}~\bibnamefont {Imada}},\ }\href
  {\doibase 10.1143/JPSJ.77.093711} {\bibfield  {journal} {\bibinfo  {journal}
  {Journal of the Physical Society of Japan}\ }\textbf {\bibinfo {volume}
  {77}},\ \bibinfo {pages} {093711} (\bibinfo {year} {2008})},\ \Eprint
  {http://arxiv.org/abs/https://doi.org/10.1143/JPSJ.77.093711}
  {https://doi.org/10.1143/JPSJ.77.093711} \BibitemShut {NoStop}%
\bibitem [{\citenamefont {Miyake}\ \emph {et~al.}(2008)\citenamefont {Miyake},
  \citenamefont {Pourovskii}, \citenamefont {Vildosola}, \citenamefont
  {Biermann},\ and\ \citenamefont {Georges}}]{cRPA_Miyake2}%
  \BibitemOpen
  \bibfield  {author} {\bibinfo {author} {\bibfnamefont {T.}~\bibnamefont
  {Miyake}}, \bibinfo {author} {\bibfnamefont {L.}~\bibnamefont {Pourovskii}},
  \bibinfo {author} {\bibfnamefont {V.}~\bibnamefont {Vildosola}}, \bibinfo
  {author} {\bibfnamefont {S.}~\bibnamefont {Biermann}}, \ and\ \bibinfo
  {author} {\bibfnamefont {A.}~\bibnamefont {Georges}},\ }\href {\doibase
  10.1143/JPSJS.77SC.99} {\bibfield  {journal} {\bibinfo  {journal} {Journal of
  the Physical Society of Japan}\ }\textbf {\bibinfo {volume} {77}},\ \bibinfo
  {pages} {99} (\bibinfo {year} {2008})},\ \Eprint
  {http://arxiv.org/abs/https://doi.org/10.1143/JPSJS.77SC.99}
  {https://doi.org/10.1143/JPSJS.77SC.99} \BibitemShut {NoStop}%
\bibitem [{\citenamefont {Miyake}\ \emph {et~al.}(2010)\citenamefont {Miyake},
  \citenamefont {Nakamura}, \citenamefont {Arita},\ and\ \citenamefont
  {Imada}}]{cRPA_ImadaMiyake}%
  \BibitemOpen
  \bibfield  {author} {\bibinfo {author} {\bibfnamefont {T.}~\bibnamefont
  {Miyake}}, \bibinfo {author} {\bibfnamefont {K.}~\bibnamefont {Nakamura}},
  \bibinfo {author} {\bibfnamefont {R.}~\bibnamefont {Arita}}, \ and\ \bibinfo
  {author} {\bibfnamefont {M.}~\bibnamefont {Imada}},\ }\href {\doibase
  10.1143/JPSJ.79.044705} {\bibfield  {journal} {\bibinfo  {journal} {Journal
  of the Physical Society of Japan}\ }\textbf {\bibinfo {volume} {79}},\
  \bibinfo {pages} {044705} (\bibinfo {year} {2010})},\ \Eprint
  {http://arxiv.org/abs/https://doi.org/10.1143/JPSJ.79.044705}
  {https://doi.org/10.1143/JPSJ.79.044705} \BibitemShut {NoStop}%
\bibitem [{\citenamefont {Miyake}\ \emph {et~al.}(2009)\citenamefont {Miyake},
  \citenamefont {Aryasetiawan},\ and\ \citenamefont {Imada}}]{cRPA_Miyake3}%
  \BibitemOpen
  \bibfield  {author} {\bibinfo {author} {\bibfnamefont {T.}~\bibnamefont
  {Miyake}}, \bibinfo {author} {\bibfnamefont {F.}~\bibnamefont
  {Aryasetiawan}}, \ and\ \bibinfo {author} {\bibfnamefont {M.}~\bibnamefont
  {Imada}},\ }\href {\doibase 10.1103/PhysRevB.80.155134} {\bibfield  {journal}
  {\bibinfo  {journal} {Phys. Rev. B}\ }\textbf {\bibinfo {volume} {80}},\
  \bibinfo {pages} {155134} (\bibinfo {year} {2009})}\BibitemShut {NoStop}%
\bibitem [{\citenamefont {Gunnarsson}\ and\ \citenamefont
  {Sch\"onhammer}(1983)}]{gunnarsson_schoenhammer_dp}%
  \BibitemOpen
  \bibfield  {author} {\bibinfo {author} {\bibfnamefont {O.}~\bibnamefont
  {Gunnarsson}}\ and\ \bibinfo {author} {\bibfnamefont {K.}~\bibnamefont
  {Sch\"onhammer}},\ }\href {\doibase 10.1103/PhysRevB.28.4315} {\bibfield
  {journal} {\bibinfo  {journal} {Phys. Rev. B}\ }\textbf {\bibinfo {volume}
  {28}},\ \bibinfo {pages} {4315} (\bibinfo {year} {1983})}\BibitemShut
  {NoStop}%
\bibitem [{\citenamefont {S\'emon}(2014)}]{PSemon_thesis}%
  \BibitemOpen
  \bibfield  {author} {\bibinfo {author} {\bibfnamefont {P.}~\bibnamefont
  {S\'emon}},\ }\emph {\bibinfo {title} {Continuous-Time Quantum Monte Carlo
  Impurity Solvers: Improvements and Applications}},\ \href@noop {} {Ph.D.
  thesis},\ \bibinfo  {school} {Facult\'e des sciences universit\'e de
  sherbrooke} (\bibinfo {year} {2014})\BibitemShut {NoStop}%
\bibitem [{\citenamefont {M{\"u}ller-Hartmann}(1989)}]{MuellerHartmann_1989}%
  \BibitemOpen
  \bibfield  {author} {\bibinfo {author} {\bibfnamefont {E.}~\bibnamefont
  {M{\"u}ller-Hartmann}},\ }\href {\doibase 10.1007/BF01311397} {\bibfield
  {journal} {\bibinfo  {journal} {Zeitschrift f{\"u}r Physik B Condensed
  Matter}\ }\textbf {\bibinfo {volume} {74}},\ \bibinfo {pages} {507} (\bibinfo
  {year} {1989})}\BibitemShut {NoStop}%
\bibitem [{\citenamefont {Amadon}\ \emph {et~al.}(2014)\citenamefont {Amadon},
  \citenamefont {Applencourt},\ and\ \citenamefont {Bruneval}}]{cRPAUO2_CE}%
  \BibitemOpen
  \bibfield  {author} {\bibinfo {author} {\bibfnamefont {B.}~\bibnamefont
  {Amadon}}, \bibinfo {author} {\bibfnamefont {T.}~\bibnamefont {Applencourt}},
  \ and\ \bibinfo {author} {\bibfnamefont {F.}~\bibnamefont {Bruneval}},\
  }\href {\doibase 10.1103/PhysRevB.89.125110} {\bibfield  {journal} {\bibinfo
  {journal} {Phys. Rev. B}\ }\textbf {\bibinfo {volume} {89}},\ \bibinfo
  {pages} {125110} (\bibinfo {year} {2014})}\BibitemShut {NoStop}%
\bibitem [{\citenamefont {Parcollet}\ \emph {et~al.}(1998)\citenamefont
  {Parcollet}, \citenamefont {Georges}, \citenamefont {Kotliar},\ and\
  \citenamefont {Sengupta}}]{Parcollet_meanfield}%
  \BibitemOpen
  \bibfield  {author} {\bibinfo {author} {\bibfnamefont {O.}~\bibnamefont
  {Parcollet}}, \bibinfo {author} {\bibfnamefont {A.}~\bibnamefont {Georges}},
  \bibinfo {author} {\bibfnamefont {G.}~\bibnamefont {Kotliar}}, \ and\
  \bibinfo {author} {\bibfnamefont {A.}~\bibnamefont {Sengupta}},\ }\href
  {\doibase 10.1103/PhysRevB.58.3794} {\bibfield  {journal} {\bibinfo
  {journal} {Phys. Rev. B}\ }\textbf {\bibinfo {volume} {58}},\ \bibinfo
  {pages} {3794} (\bibinfo {year} {1998})}\BibitemShut {NoStop}%
\end{thebibliography}
%
\end{document}